# Survey of Important Issues in UAV Communication Networks

Lav Gupta*, *Senior Member IEEE*, Raj Jain, *Fellow, IEEE,* and Gabor Vaszkun

*Abstract*—Unmanned Aerial Vehicles (UAVs) have enormous potential in the public and civil domains. These are particularly useful in applications where human lives would otherwise be endangered. Multi-UAV systems can collaboratively complete missions more efficiently and economically as compared to single UAV systems. However, there are many issues to be resolved before effective use of UAVs can be made to provide stable and reliable context-specific networks. Much of the work carried out in the areas of Mobile Ad Hoc Networks (MANETs), and Vehicular Ad Hoc Networks (VANETs) does not address the unique characteristics of the UAV networks. UAV networks may vary from slow dynamic to dynamic; have intermittent links and fluid topology. While it is believed that ad hoc mesh network would be most suitable for UAV networks yet the architecture of multi-UAV networks has been an understudied area. Software Defined Networking (SDN) could facilitate flexible deployment and management of new services and help reduce cost, increase security and availability in networks. Routing demands of UAV networks go beyond the needs of MANETS and VANETS. Protocols are required that would adapt to high mobility, dynamic topology, intermittent links, power constraints and changing link quality. UAVs may fail and the network may get partitioned making delay and disruption tolerance an important design consideration. Limited life of the node and dynamicity of the network leads to the requirement of seamless handovers where researchers are looking at the work done in the areas of MANETs and VANETs, but the jury is still out. As energy supply on UAVs is limited, protocols in various layers should contribute towards greening of the network. This article surveys the work done towards all of these outstanding issues, relating to this new class of networks, so as to spur further research in these areas.

## I. INTRODUCTION

### A. *The growing importance of UAV networks*

Unmanned Aerial Vehicles (UAVs) are an emerging technology that can be harnessed for military, public and civil applications. Military use of UAVs is more than 25 years old primarily consisting of border surveillance, reconnaissance and strike. Public use is by the public agencies such as police, public safety and transportation management. UAVs can provide timely disaster warnings and assist in speeding up rescue and recovery operations when the public communication network gets crippled. They can carry medical supplies to areas rendered inaccessible. In situations like poisonous gas infiltration, wildfires and wild animal tracking UAVs could be used to quickly envelope a large area without risking the safety of the personnel involved.

UAVs come in various sizes. Large UAVs may be used singly in missions while small ones may be used in formations or swarms. The latter ones are proving to be quite useful in civilian applications. As

> At a recent show in Decatur, Illinois, on farming applications of small UAVs drew 1400 attendees from 33 states and 6 countries. According to Stu Ellis, an organizer, "You could spend four to five hours walking an 80-acre soybean field," and noted that the same ground could be covered in half an hour or less by a drone. However, in its June 2014 notice, FAA has made it clear that drones that are not being used by hobbyists or any commercial use must have prior FAA approval. The agency specifically mentions farming – along with photography and delivery services – as types of businesses subject to regulation. Meanwhile, Association for Unmanned Vehicles Systems International has projected an $82 billion economic impact for the period 2015-2025. (Adapted from St Louis Post dated 07/27/2014)

described by Daniel and Wietfeld in [1] they are likely to become invaluable inclusions in the operations of police departments, fire brigades and other homeland security organizations in the near future. Besides, advances in electronics and sensor technology have widened the scope of UAV network applications [2] to include applications as diverse as traffic monitoring, wind estimation and remote sensing [3].

In this context it would be relevant to mention that the current FAA guidelines allow a government public safety agency to operate unmanned aircraft weighing 4.4 pounds or less, within the line of sight of the operator; less than 400 feet above the ground; during daylight conditions; within Class G airspace; and outside of 5 statute miles from any airport, heliport, seaplane base, spaceport, or other location. For model aircrafts FAA guidance does not address size of the model aircraft. The guidelines of Federal Aviation Authority [4] say that model aircraft flights should be kept below 400 feet above ground level (AGL), should be flown a sufficient distance from populated areas and full scale aircraft, and are not for business

*Formal corresponding author
The manuscript was submitted on 14th October 2014.
Lav Gupta is pursuing doctoral program in Computer Science & Engineering at Washington University in St Louis, MO 63130 USA (email: lavgupta@wustl.edu).
Raj Jain is Professor of Computer Science & Engineering at Washington University in St Louis, MO 63130, USA (email: jain@cse.wustl.edu).
Gabor Vaszkun is with Ericsson Hungary (email: vaszkun@gmail.com)

purposes.

*B. The challenges of UAV networks*

Promising though it may be, this area is relatively new and less explored. There are many issues to resolve before effective use of UAVs can be made to provide stable and reliable context-specific networks. As we shall see later, while it offers the promise of improved capability and capacity, establishing and maintaining efficient communications among the UAVs is challenging.

All the constituents of the UAV communication networks pose challenging issues that need resolution. Unlike many other wireless networks, the topology of UAV networks remains fluid with the number of nodes and links changing and also the relative positions of the nodes altering. UAVs may move with varying speeds depending on the application, this would cause the links to be established intermittently. What challenges would such a behavior pose? Firstly, some aspects of the architectural design would not be intuitive. The fluid topology, the vanishing nodes and finicky links would all challenge the designer to go beyond the normal ad hoc mesh networks. Second, the routing protocol cannot be a simple implementation of a proactive or a reactive scheme. The inter-UAV backbone has to repeatedly reorganize itself when UAVs fail. In some cases the network may get partitioned. The challenge would then be to route the packet from a source to a destination while optimizing the chosen metric. The third challenge would be to maintain users' sessions by transferring them seamlessly from an out of service UAV to an active UAV. Lastly, there need to be ways of conserving energy of power starved UAVs for increasing the life of the network. In the next section we bring out all of these issues in more detail.

*C. Motivation and key Issues*

The area of UAV networks is challenging to researchers because of the outstanding issues that provide motivation for research. In mobile and vehicular networks the nodes join and dissociate from the network frequently and, therefore, ad hoc networks have been found to be suitable in most situations. In addition, for quick and reliable communication between nodes, mesh network topology is quite appropriate. Does this apply to the UAV networks as well? In UAV networks, the nodes could almost be static and hovering over the area of operation or scouting around at a rapid pace. Nodes could die out for many reasons and may be replaced by new ones. Some similarities encourage researchers to explore the applicability of the work done for Mobile Ad hoc Networks (MANETs) and Vehicular Ad hoc Networks (VANETs), but works in these areas do not fully address the unique characteristics of the UAV networks. Table I gives important characteristics of MANETs, VANETs and UAV networks which bring out similarities and dissimilarities among them. We shall characterize UAV networks in more depth in section II(C).

TABLE I
COMPARATIVE DESCRIPTION OF DIFFERENT AD-HOC NETWORKS

| | MANET | VANET | UAV Networks |
|---|---|---|---|
| Description | Mobile wireless nodes connect with other nodes within communication range in an ad-hoc manner (No centralized infrastructure required) | Ad-hoc networks in which vehicles are the mobile nodes. Communication is among vehicles and between vehicles and road side units | Ad-hoc or infrastructure based networks of airborne nodes. Communication among UAVs and with the control station |
| Mobility | Slow. Typical speeds 2 m/sec. Random movement. Varying density, higher at some popular places | High-speed, typically 20-30 m/s on highways, 6-10 m/s in urban areas. Predictable, limited by road layout, traffic and traffic rules | Speeds from 0 to typically as high as 100 m/s. Movement could be in 2 or 3 dimensions, usually controlled according to mission. |
| Topology | Random, ad-hoc | Star with roadside infrastructure and ad-hoc among vehicles | Star with control center, ad-hoc/mesh among UAVs. |
| Topology Changes | Dynamic - nodes join and leave unpredictably. Network prone to partitioning. | More dynamic than MANETs. Movement linear. Partitioning common. | Stationary, slow or fast. May be flown in controlled swarms. Network prone to partitioning |
| Energy Constraints | Most nodes are battery powered so energy needs to be conserved. | Devices may be car battery powered or own battery powered. | Small UAVs are energy constrained. Batteries affect weight and flying time |
| Typical use cases in public and civil domains | • Information distribution (emergencies, advertising, shopping, events)<br>• Internet hot spots | • Traffic & weather info, emergency warnings, location based services<br>• Infotainment | • Rescue operations<br>• Agriculture-crop survey<br>• Wildlife search<br>• Oil rig surveillance |

It can be observed from Table 1 and some of the references [3] [123] that there are many aspects of UAV networks that set them apart from mobile ad-hoc and vehicular ad-hoc networks. The important differences are described here:

The mobility models, like random walk, that have been used to describe the behavior of nodes in mobile ad hoc networks and street random walk or Manhattan models for vehicular ad hoc networks are not quite suitable for UAV networks. The UAVs could move randomly or in organized swarms not only in two but also in three dimensions with rapid change in position. The vehicular nodes are constrained to travel on roads and only in two dimensions.

Changes in topology could be much more frequent in UAV networks. Relative positions of UAVs may change; some UAVs may lose all their power and may need to be brought down for recharging; UAVs may malfunction and be out of the network; the links may form and vanish because of the changing positions of the nodes. In many applications density of nodes may not be high and the network may partition frequently. Vehicular networks have roadside infrastructure to support communication among vehicles. The network being fluid, the mobile stations that are locked on to a particular UAV for communication access would need to be transferred to another UAV seamlessly, without disrupting user sessions. The research community is still trying to figure out the most effective routing protocol as well as the seamless handover procedure [7].

Energy constraints are much greater in small UAV networks. While in vehicular networks power could be drawn from the car battery which gets charged while the car is in motion. Even the mobile ad hoc networks would typically have nodes (smartphones, laptops) with power sources that last a few hours. Small UAVs may typically have enough power for a flight of thirty minutes! On one hand the transmitted signal would be of lower power and on the other the links would be intermittent owing to power-drained UAVs drifting away or dying. Dynamicity of nodes would force the network to organize and re-organize frequently. This gives rise to unique routing requirements. The routing protocols need to use energy efficiently so as to prolong the stability of the UAV network.

The UAV networks would usually be deployed in dire circumstances and the network may get frequently partitioned, sometimes for long durations. In such cases traditional solutions do not guarantee connectivity. It has been suggested that these networks be made delay and disruption tolerant by incorporating store-carry-forward capabilities. If we presume that each node has a global knowledge of the network we could apply deterministic protocols but this may not always be a correct assumption. On the other hand if we presume random behavior of nodes we have to face scalability issues.

A multi-UAV network, which is fully autonomous, requires a robust inter-UAV network with UAVs to cooperate in keeping the network organized even in the event of link or node failure. The UAV networks would require changes at the MAC and network layers, have self-organizing capabilities, be tolerant to delays, have a more flexible and automated control through SDN and employ energy saving mechanisms at various layers [5], [6].

That these issues need thorough attention is corroborated in a well-detailed survey on UAV based flying ad hoc network [3]. Authors of this survey have tried to put together many of the issues and developments in this area. However, there are issues of self-organization, disruption tolerance, SDN control, seamless handover and energy efficiency that we believe, would be extremely important for building successful UAV networks, have not been the foci of the study. Our survey attempts to primarily take up these issues to bring out the current status and possible directions.

The organization of rest of the paper is as follows: In Section II, we attempt to categorize UAV networks and examine important characteristics like topology, control, and client server behavior. We also see the important aspects of self-organization and automated operations through software defined networking (SDN) that will help in identifying the work required in this area. In Section III, we discuss requirements from the routing protocols peculiar to UAV networks and the need for disruption tolerant networking. In Section IV, we see the importance of seamless handover and the need for new research in this area. Finally, in Section V, the protocols used at various layers for energy conservation are discussed.

## II. CHARACTERIZING THE UAV NETWORK

It is important to characterize a network to understand its nature, constraints and possibilities. How fast does the topology change with time? How frequently does the network get partitioned as the nodes fail or move away? How can the network life be increased? What type of architecture would be more suitable? Does it require self-organizing, self-healing capabilities? Which protocols can be run at different layers? Does it support addition and removal of nodes dynamically? Are the links intermittent and what is their quality? In this section we look at the characteristics that forms the common thread in various works and the direction in which the research is headed. Subsection A contains characteristics of multi-UAV systems and their advantages over single UAV systems. Subsection B discusses important features that set them apart from other ad hoc networks and also distinguishes them from each other. Subsection C provides categorization of UAV networks based on some of the important characteristics discussed. Subsection D brings forth the self-organizing behavior of UAV networks. Lastly, Subsection E deals with less uncovered aspect of use of SDN for centralizing and automating control of UAV networks.

### A. Multi-UAV network

Early uses of UAV were characterized by use of a single large UAV for a task. In these systems the UAV based communication network, therefore, consisted of just one aerial node and one or more ground nodes. Today most public and civil applications can be carried out more efficiently with multi UAV systems. In a multi-UAV system, the UAVs are smaller and less expensive and work in a coordinated manner. In most multi-UAV systems, the communication network, proving communication among UAVs and between the UAVs and the ground nodes, becomes an important constituent. These UAVs can be configured to provide services co-operatively and extend the network coverage by acting as relays. The degree of mobility of UAVs depends on the application. For instance, in providing communication over an earthquake struck area the UAVs would hover over the area of operation and the links would be slow dynamic. As opposed to this, applications like agriculture or forest surveillance require the UAVs to move across a large area and links frequently break and reestablish. The dynamic nature of the network configuration and links is apparent from the fact that the UAVs may go out of service periodically due to malfunction or battery drainage. This is true also for UAVs that need to hover over an area for relatively long periods. New UAVs have to be launched to take their places. Sometimes some of the UAVs may be taken out of service to conserve power for a more appropriate time. It would, therefore, be a requirement that in all such cases the links should automatically reconfigure themselves. Though advantageous in many respects, multi UAV systems, add complexities to the UAV communication network.

Some of the key advantages of multi-UAV systems are reliability and survivability through redundancy. In a multi-UAV system, failure of a single UAV causes the network to re-organize and maintain communication through other nodes. This would not be possible in a single UAV system. However, to reap the real benefits of the multiple UAVs working in collaboration, the protocols deployed need to take care of the issues like changing topology, mobility and power constraints. In terms of communication needs, single UAV systems would have to maintain links with the control station(s), base stations, servers and also provide access functionality. This puts a heavy constraint on the limited battery power and bandwidth. In a multi-UAV system, only one or two UAVs may connect to control and servers and feed the other UAVs. This way most UAVs just have to sustain the mesh structure and can easily offer access functions for calls, video and data. Multi-UAV systems also turn out to be less expensive to acquire, maintain and operate than their larger counterparts. As shown through their experiments by Mergenthaler et al. in [8], adding more UAVs to the network can relatively easily extend communication umbrella provided by a multi-UAV system. Missions are generally completed more speedily, efficiently and at lower cost with small UAV systems as compared to a single UAV system [9]. In their work on PRoPHET-based routing protocol [10] the authors explain how in opportunistic networks multi-UAVs find a path even if two end points are not directly connected leading to completion of missions. In their work on Multi-UAV cooperative search in [11] the authors describe how multi-UAV systems complete searches faster and are robust to loss of some UAVs. Advantages of multi-UAV networks are leading to their increasing use in civilian applications [1]. In this paper we focus on the UAV networks that use multiple small UAVs to form an unmanned aerial system (UAS). Table II gives a comparison of single and multi-UAV systems.

TABLE II
COMPARISON OF SINGLE AND MULTI UAV SYSTEMS

| FEATURE | SINGLE UAV SYSTEM | MULTI UAV SYSTEM |
| --- | --- | --- |
| Impact of failure | High, mission fails | Low, system reconfigures |
| Scalability | Limited | High |
| Survivability | Poor | High |
| Speed of mission | Slow | Fast |
| Cost | Medium | Low |
| Bandwidth required | High | Medium |
| Antenna | Omni-directional | Directional |
| Complexity of control | Low | High |
| Failure to coordinate | Low | Present |

An issue that is important in the context of multi-UAV systems, but not a subject of this study is coordination and control for effective task planning with multiple UAVs. An efficient algorithm is necessary to maneuver each UAV so that the whole system can produce complex, adaptable and flexible team behavior. The task-planning problem for UAV networks with connectivity constraint involves a number of parameters and interaction of dynamic variables. An algorithm has been proposed for such a situation [11]. There is also an algorithm for distributed intelligent agent systems in which agents autonomously coordinate, cooperate, negotiate, make decisions and take actions to meet the objectives of a particular task. The connectivity constrained problem is NP-hard and a polynomial time heuristic has been proposed in the literature [13]. This is a challenging field but we would restrict our attention to UAV based communication networks.

*B. Features of the UAV networks*

Developing a fully autonomous and cooperative multi-UAV system requires robust inter-UAV communication. We do not have enough research to say with conviction what design would work best. There are a number of aspects of the UAV networks that are not precisely defined and a clarification of these would help in characterizing the UAV networks:

*1) Infrastructure-based or ad hoc?*

Most of the available literature treats UAV networks as ad hoc networks. Research on MANETs and VANETs are often cited with reference to UAV networks but they do not completely address the unique characteristics of the UAV networks. Depending on the application, the UAV network could have stationary, slow moving or highly mobile nodes. Many applications require UAV nodes to act as base stations in the sky to provide communication coverage to an area. Thus, unlike MANET and VANET ad hoc networks, the UAV networks could behave more like infrastructure-based networks for these applications. These would have UAVs communicating with each other and also with the control center. Such a network would resemble the fixed wireless network with UAVs as base stations except that they are aerial. There is, however, a class of applications where the nodes would be highly mobile and would communicate, cooperate and establish the network dynamically in an ad hoc manner. In such a case the topology may be determined, and the nodes involved in forwarding data decided, dynamically. There are many issues that affect both UAV infrastructure based and UAV ad hoc networks. For example, replacing nodes by new nodes when they fail or their power gets exhausted.

*2) Server or client?*

Another point of distinction is whether the node acts as a server or a client. In vehicular networks they are usually clients, in mobile ad hoc networks most of the time they would be clients and may also provide forwarding services to other clients' data. In UAV networks, the UAV nodes are usually servers, either routing packets for clients or relaying sensor data to control centers.

*3) Star or Mesh?*

Architecture of UAV networks for communication applications is an understudied area. The simplest configuration is a single UAV connected to a ground based command and control center. In a multi-UAV setting, the common topologies that can be realized are star, multi-star, mesh and hierarchical mesh. In the case of star topology all UAVs would be connected directly to one or more ground nodes and all communication among UAVs would be routed through the ground nodes. This may result in blockage of links, higher latency and requirement of more expensive high bandwidth downlinks. In addition, as the nodes are mobile, steerable antennas may be required to keep oriented towards ground node [15]. The multi-star topology is quite similar except the UAVs would form multiple stars and one node from each group connects to the ground station. Figures 1a and 1b shows star and multi-star configurations.

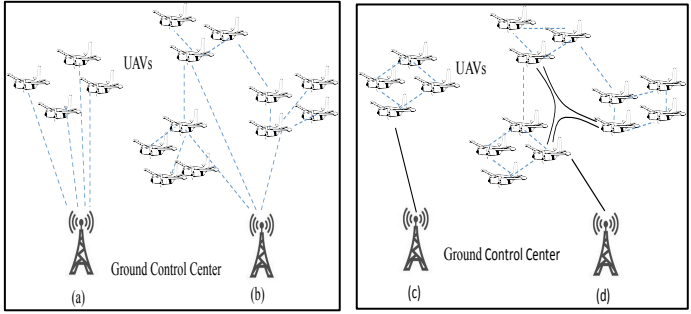

Fig. 1a) Star Configuration b) Multi-star Configuration c) Flat Mesh Network d) Hierarchical Mesh Network

Star configurations suffer from high latency as the downlink length is longer than inter-UAV distance and all communication must pass through the ground control center. Also, if the ground center fails there is no inter-UAV communication. In most civilian applications, however, normal operation does not require communication among UAVs to be routed through the ground node. An architecture that supports this would result in reduced downlink bandwidth requirement and improved latency because of shorter links among UAVs. In case of mesh networks the UAVs are interconnected and a small number of UAVs may connect to the control center [16]. Figures 1c and 1d show flat and hierarchical mesh networks.

Some authors believe that conventional network technologies cannot meet the needs of UAV networks. Related literature points to the applicability of mesh networks for civilian applications [15, [9]. There are usually multiple links on one or more radios, interference between channels, changes in transmitted power due to power constrains, changes in number of nodes, changes in topology, terrain and weather effects. In ad hoc networks the nodes may move away, formations may break, and therefore, the links may be intermittent. Wireless mesh networks, adapted suitably, may take care of some of these problems. To tackle these issues the network has to be self-healing with continuous connection and reconfiguration around a broken path.

As compared to star networks, mesh networks are flexible, reliable and offer better performance characteristics. In a wireless mesh network, nodes are interconnected and can usually communicate directly on more than one link. A packet can pass through intermediate nodes and find its way from any source to any destination in multiple hops. Fully connected wireless networks have the advantages of security and reliability. Such a network can use routing or flooding technique to send messages. The routing protocol should ensure delivery of packets from source to destination through intermediate nodes. There are multiple routes and the routing protocol should select the one that meet the given objectives. The routing devices can organize themselves to create an ad hoc backbone mesh infrastructure that can carry users' messages over the coverage area through multiple hops. Moreover, they can also route packets originated from the command and control center and directed to emergency operators or people and vice-versa. The control center can process data to extract information for the support of decisions during emergency [17]. Due to unique features of UAV nodes described above, sometimes the existing networks routing algorithms, which have been designed for mobile ad hoc networks (MANETs), such as BABEL or the Optimized Link-State Routing (OLSR) protocol, fail to provide reliable communications [8], [14]. We shall provide more details about routing in the next section. The major differences between star and mesh networks are given in Table III.

TABLE III
COMPARISON OF STAR AND MESH NETWORKS PROPERTIES

| Star Network | Mesh Network |
| --- | --- |
| Point-to-point | Multi-point to multi-point |
| Central control point present | Infrastructure based may have a control center, Ad hoc has no central control center |
| Infrastructure based | Infrastructure based or Ad hoc |
| Not self configuring | Self configuring |
| Single hop from node to central point | Multi-hop communication |
| Devices cannot move freely | In ad hoc devices are autonomous and free to move. In infrastructure based movement is restricted around the control center |
| Links between nodes and central points are configured | Inter node links are intermittent |
| Nodes communicated through central controller | Nodes relay traffic for other nodes |
| Scalable | Not scalable |

*4) Delay and Disruptions prone networks*

All wireless mobile networks are prone to link disruptions. The UAV networks are no exception. The extent of disruption depends on how mobile the UAVs are, the power transmitted, inter-UAV distances and extraneous noise. In the applications where UAVs provide communication coverage to an area, the UAVs are hovering and, therefore, probability of disruptions would be low. On the other hand, in applications requiring fast UAV mobility, there is a higher likelihood of disruptions. Delays in transmitting data could be because of poor link quality or because one or more UAV nodes storing the data because of end-to-end path not being available. We will see more details of this in Section III.

*C. Categorization of UAV Networks*

We are now ready to categorize the UAV networks. In Subsection 2.2 we have seen a number of features that could form the basis of this categorization. So how do we categorize UAV networks, the applications of which require varying degree of node mobility, different network architectures, routing and control? In the ultimate analysis it is the applications that would be important. Therefore, considering distinct applications, and then deciding the properties that would needed to make these applications possible, could lead us to a generic categorization as given in Table IV.

Like Internet delivery there are many applications in which UAV assisted wireless infrastructure needs to be established for providing area wide coverage. Disaster struck areas, remote villages or deep sea oil rigs would all require quick deployment of a UAV based network that could provide voice, video and data services. In these applications the UAVs hover over an area and are virtually stationary. These can be considered to be cellular towers or wireless access points in the sky. In contrast, sensing applications like detection of forest fires or survey of crops would require mobile UAVs. There are other applications, especially military, where fast moving UAVs would be required to make forays into enemy territory. We would focus here largely on the first category of applications and to some extent on the sensing applications. When UAVs are used to create wireless communication infrastructure then depending on the application, all UAVs could be directly under control of the ground control center or they could form a wireless mesh network with one or two UAVs communicating with the control center. In these applications the UAVs act as servers for routing users' communication and control information. This is distinct, for example, from the cases where UAVs carrying sensors are used to collect information or those sent out for carrying out attack. In these cases, UAVs act as clients. Delay or disruption probability in the Internet delivery class of applications is much less as compared to other applications because the UAVs are stationary and formations are easily coordinated. When a UAV fails, the network would be expected to reconfigure and the ongoing sessions would have to be seamlessly transferred to other UAVs. In other applications where nodes are more dynamic, the links may function intermittently and special routing, reconfiguration, and disruption handling features may be required. In the following sections we shall see in more detail the state of research in these areas.

TABLE IV
UAV CATEGORIZATION

| Property | Internet Delivery | Sensing | Attack |
|---|---|---|---|
| Other general applications in the category | Disaster communication, Oil exploration Remote health | Reconnaissance, search, detecting forest fires, tracking wild animals | War: Multi-UAV attack |
| UAV Position during communication | Fixed position | Slow change, coordinated movement | Frequent change |
| Mobility speed of UAV during communication | ~0 miles/hour | <10 miles/hour | >10 miles/hour |
| Network | Infrastructure based, base station in the sky | Infrastructure based | Infrastructure based/Infrastructure less, Ad-hoc |
| Topology | Star/Mesh | Mesh | Mesh |
| Control (communication) | Centralized (position control based) | Centralized ( task control based) | Distributed (task control based) Individuals controlling each UAV |
| UAV as Client or Server | Server (routes communication and control) | Server (when receiving from sensors) / client(when carrying sensors | Server (delivering info to formations) /client (for attack) |
| Routing | Through server (control from central, data through central or among UAVs) | Central or mesh (control from central data to central, also data among UAVs) | Mesh routing (control from central, data among UAVs) |
| Delay/disruption (because of node and link failures) | Low probability (p<0.1) | Medium probability (0.1<=p<0.5) | High probability (p>=0.5) |
| Type of Communication C=Client, I=Infrastructure | U2U, U2I, C2U | U2U (if buffer node), U2I, U2C | U2I (commands), U2U |
| Control( path control, position) | Remote – position control | Remote – position control, path control | Remote (less likely) Auto (observe, orient, decide, act) Path Control – real-time multi UAV coordination, collision avoidance |

*D. Self-organization in networks*

One reason why mesh networks are considered suitable for UAV based networks is because of their self-forming and reorganization features. Once the nodes have been configured and activated they form mesh structure automatically or with guidance from the control center. When this happens, the network becomes resilient to failure of one or more nodes. There is inherent fault tolerance in mesh networks. When a node fails, the rest of the nodes reconfigure the network among them. By the same token a new node can be easily introduced. Support for ad hoc networking, self-forming, self-healing and self-organization enhance the performance of wireless mesh networks, make them easily deployable and fault tolerant [18] [19].

Studies on self-organization in the context of sensor networks and wireless ad hoc networks can help in understanding the requirements of UAV networks. Self-organization consists of the following steps: When a node fails or a new node appears, its neighbor(s) finds out about the available nodes through the neighbor discovery process. Changes in the network, in the form of removal or addition of devices in the network, cause a number of nodes to exchange messages for re-organization. This could cause collisions while accessing the medium and impact network performance. Medium access control deals with control of access to the medium and minimizes errors due to collisions. The next step involves establishment of connectivity between nodes during self-organization through local connectivity and path establishment. Once the connectivity has been established, the service recovery management process carries out network disruption avoidance and recovery from local failures. Finally, energy management carries out load balancing of data forwarding responsibilities in the self-organized network and also the processes involved in reducing energy consumption in the battery operated devices. The goal in UAV networks would then be to ensure connectivity of all active network nodes so that the mesh network is maintained through multi-hop communication to provide best access to the users [20].

Wireless mesh networks are prone to link failure due to interference, mobility or bandwidth demand. This will lead to degradation of network performance but can be effectively tackled by making the network reconfigurable. The nodes monitor their links and any failures trigger the reconfiguration process. The autonomous network reconfiguration system requires computational overhead and reasonable bandwidth. There are times when some of the UAVs could go out of service because of battery drain or communication failure. In such cases the remaining nodes in the network re-organize and re-establish communication. While the benefits of self-organization are huge and encouraging, challenges in self-organization are bigger making this an exciting research problem [21].

*E. SDN-automating UAV network control*

The UAV networks are limited in communication resources. Nodes are non-permanent, connectivity is intermittent and channels may be impaired. This translates into utilization challenges in planning and allocation of resources. Different networks utilize different routing protocols and consequently even the nodes that use the same access technology may not work in another network because of the differences in higher layers of the protocol stack. For instance vehicular networks use IEEE 802.11p Wireless Access Vehicular Environment (WAVE) to support Intelligent Transportation Systems (ITS) applications. For wireless mesh networks, the IEEE 802.11s amendment has been standardized for self-configuring multi-hop technologies. But in both environments there is no consensus on the routing protocols to be used, as well as on most of the network management operations to be performed. As a consequence, nodes using a particular access technology in a network may not operate in another network with same access but different higher layer protocols.

The above problems could be tackled by building in the capability of defining the protocol stack in software. This way the UAVs could be programmed to flexibly work in different environments. However, this is not the only reason why software control of the network is desirable. There are a number of other requirements arising in case of networks like MANET, VANET and UAV networks. They need to support dynamic nodes and frequent changes in the topology. Nodes may fail, for example, because of battery drainage and need to be replaced by new nodes. The links are intermittent and need to be accordingly dealt with. SDN provides a way to programmatically control networks, making it easier to deploy and manage new applications and services, as well as tune network policy and performance [22], [23].

Deployment of SDN has been extensive in fixed infrastructure-based networks. However, much of this deployment has been in datacenters, as it was believed that SDN was suitable for networks with centralized control and wireless ad hoc mesh networks were decentralized. Separation of forwarding devices and the controller also raised the concern for security, balance of control and flexibility. There are several policy issues relating to balance of control and cooperation between controllers that need to be addressed. Because of the benefits it is expected to provide, increasing interest is being shown by the academia and the industry in application of SDN in dynamic mobile wireless environments. In networks like VANETs, the use of SDN can help in path selection and channel selection. This helps in reducing interference, improving usage of wireless resources including channels and routing in multi-hop mesh networks. Despite increasing interest there is no clear and comprehensive understanding of what are the advantages of SDN in infrastructure-less wireless networking scenarios and how the SDN concept should be expanded to suit the characteristics of wireless and mobile communications [22], [24].

One of the commonly used protocols for implementing SDN in wireless networks is OpenFlow. OpenFlow is claimed to deliver substantive advantages for mobile and wireless networks. It helps in optimizing resource usage in a dynamic environment, provides a way to automate operations, allows finer level of control and easier implementation of the global policies and faster introduction of new services [25]. OpenFlow protocol separates forwarding, and control functionalities. The OpenFlow switches are programmable and contain flow tables and protocol for communication with controller(s) [26]. Figure 2 shows separation of control and forwarding functionalities with OpenFlow interface between control and data planes.

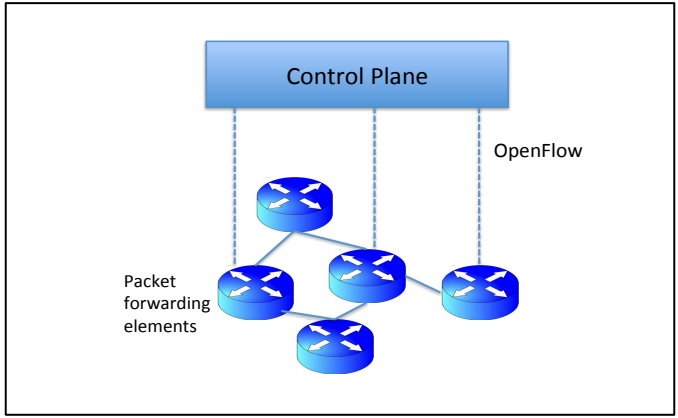

Fig. 2: Software Defined Networking elements

Such a network may be formed by mounting the data plane or the OpenFlow switches on the UAVs and the control facility on a centralized ground controller or distributed control on UAVs. The forwarding elements, which are simple OpenFlow switches, rather than IP routers, contain the flow tables that are manipulated by the controller to set the rules. The actions in the flow tables define processing that would be applied to the identified set of packets. Actions could be filtering, forwarding a particular port, header rewriting etc. Figure 3a shows how the centralized control plane base SDN can be implemented in a UAV network.

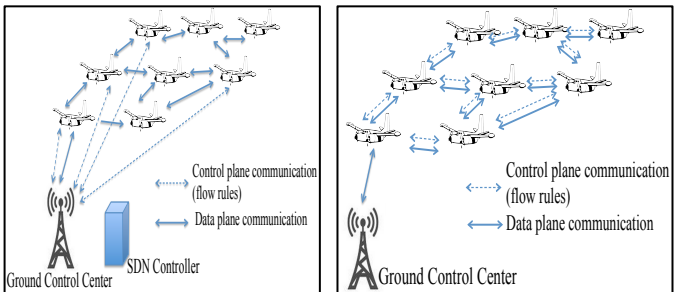

Fig. 3: a) SDN with centralized control b) SDN with distributed control

The control plane of the SDN network could be centralized, distributed or hybrid. The controller has a global view of the network and can effectively route traffic. It updates flow tables of the OpenFlow switches with matching criteria and processing actions [27]. In the centralized mode, that we saw above, the controller, that defines all the actions that wireless switches in the UAV take, has been shown to be on the ground. It could be air-borne as well. Figure 3b shows distributed control of the SDN network. In the distributed control mode the control is distributed on all the UAVs and each node controls its behavior. In the hybrid mode the controller delegates control of packet processing to the local agent and there is control traffic between all SDN elements.

Researchers at Karlstad University, Sweden have demonstrated the feasibility of integration of OpenFlow with wireless mesh networks on their KAUMesh testbed on a small scale [28]. Simulation by authors in [23] shows the comparison of SDN-based VANET routing to other traditional MANET/VANET routing protocols, including GPSR, OLSR, AODV, and DSDV. They observe that SDN-based routing outperforms the other traditional Ad hoc routing protocols in terms of packet delivery ratio at various speeds. The major reason for this is the aggregated knowledge that the SDN controller has. In [29] the authors summarize the current situation by saying that SDN offers enhanced configuration, improved performance and encourages innovation but it is still in infancy and many fundamental issues still remain not fully solved. Capabilities of SDN to fulfill various requirements of a UAV mesh network are given in Table V.

TABLE V
UAV COMMUNICATION REQUIREMENTS AND SDN CAPABILITIES

| Feature required in UAV | SDN capability |
| --- | --- |
| Support for node mobility | Reconfiguration through orchestration mechanism |
| Flexible switching and routing strategies | Flexible definition of rules based on header or payload for routing data. |
| Dealing with unreliable wireless links | Selection of paths and channels |
| Greening of network | Supports switching off devices when not in use. Supports data aggregation in the network |
| Reduce interference | Can be done through path/channel selection |

## III. ROUTING

The UAV networks constituted for various applications may vary from slow dynamic to the ones that fly at considerable speeds. The nodes may go out of service due to failure or power constraints and get replaced by new ones. In green networks, radios in the nodes may be automatically switched off for power conservations when the load is low. Link disruption may frequently occur because of the positions of UAVs and ground stations. Additionally, the links could have high bit error rates due to interference or natural conditions. The reliability requirements from the UAV networks are also diverse. For example, while sending earthquake data may require a 100% reliable transport protocol, sending pictures and video of the earthquake may be done with lower reliability but stricter delay and jitter requirements. Bandwidth requirements for voice, data and video are different. The UAV networks, therefore, have all the requirements of mobile wireless networks and more. Node mobility, network partitioning, intermittent links, limited resources and varying QoS requirements make routing in UAV a challenging research task. Several existing routing protocols have been in the zone of consideration of researchers and quite a few variations have been proposed. While newer protocols attempt to remove some shortcoming of the traditional ones, the field is still open. Subsection A discusses the important issues that routing protocols need to deal with. Applicability of various traditional protocols to UAV networks has been discussed in Subsection B. Also included in this subsection are new variants of these traditional protocols as well as some recently proposed UAV specific routing protocols. UAV networks are prone to delays and disruptions and Subsection C discusses the routing protocols that can be used in such situations.

*A. Routing issues to be resolved*

In addition to the requirements present in the generic wireless mesh networks, e.g., finding the most efficient route, allowing the network to scale, controlling latency, ensuring reliability, taking care of mobility and ensuring required QoS; routing in airborne networks requires location-awareness, energy-awareness, and increased robustness to intermittent links and changing topology. Designing the network layer for UAV networks is still one of the most challenging tasks [3].

There are some papers that have looked into the use of existing routing protocols for possible use in airborne networks. Although conventional ad hoc routing protocols are designed for mobile nodes, they are not necessarily suitable for airborne nodes because of varying requirements of dynamicity and link interruptions. Therefore, there still exists a need for a routing protocol tailored to the particular needs of airborne networks that adapts to high mobility, dynamic topology and different routing capabilities [30]. Routing protocols try to increase delivery ratio, reduce delays and resource consumption. Additionally, one has to consider issues related to scalability, Loop freedom, energy conservation, and efficient use of resources also needs to be resolved [31].

*B. Applicability of existing routing protocols*

A number of routing protocols that have been proposed for MANETs try to adapt the proactive, table based, protocols of the wired era to ad hoc wireless networks with mobile nodes. Many of these and also on-demand or reactive protocols suffer from routing overhead problems and consequently have scalability and bandwidth issues. Conditional update based protocols reduce overheads but location management remains an issue in dynamic networks like UAV networks. Some protocols that introduce the concept of cluster heads introduce performance issues and single point of failure [32]. A survey on WMN stationary or mobile nodes, points out that available MAC and routing protocols do not have enough scalability and the throughput drops significantly as the number of nodes or hops increase. It goes on to add that protocol improvement in a single layer will not solve all the problems so all existing protocols need to be enhanced or replaced by new ones for UAV Networks [33].

Due to apparent similarity of UAV networks with MANETs and VANETs researchers have studied protocols used in those environments for possible application in aerial networks. Even in these environments the search for more improved protocols is on. However, multi-UAV networks may have a number of different requirements to take care of e.g. mobility patterns and node localization, frequent node removal and addition, intermittent link management, power constraints, application areas and their QoS requirements. Due to many of these issues peculiar to UAV networks, while modifications have been proposed to MANET protocols, there is a need to develop new routing algorithms to have reliable communication among UAVs [7] and from UAVs to the control center(s). We shall discuss a number of networking protocols using the following well-known classification with a view to see their usefulness for UAV Networks: 1) Static protocols 2) Proactive protocols 3) On-demand or Reactive protocols, and 4) Hybrid protocols.

*1) Static Routing Protocols*

Static protocols have static routing tables, which are computed and loaded when the task starts. These tables cannot be updated during an operation. Because of this constraint, these systems are not fault tolerant or suitable for dynamically changing environment. Their applicability to UAV networks is limited and is only presented here for academic interest.

- One such protocol is *Load Carry and Deliver Routing (LCAD)* [34] in which the ground node passes the data to a UAV, which carries it to the destination. It aims to maximize the throughput while increasing security. Because of use of a single UAV, data delivery delays are longer in LCAD but it achieves higher throughput. LCAD can scale its throughput by using multiple UAVs to relay the information to multiple destinations. It is possible to use this for delay tolerant and latency insensitive bulk data transfer.
- *Multi Level Hierarchical Routing [MLHR]* [7] solves the scalability problems faced in large-scale Vehicular networks. These networks are normally organized as flat structures because of which performance degrades when the size increases. Organizing the network as hierarchical structure increases size and operation area. Similarly UAV networks can be grouped into clusters where only the cluster head has connections outside cluster. The cluster head disseminates data by broadcasting to other nodes in the cluster. In UAV networks frequent change of cluster head would impose large over head on the network. Figure 4 depicts this pictorially.

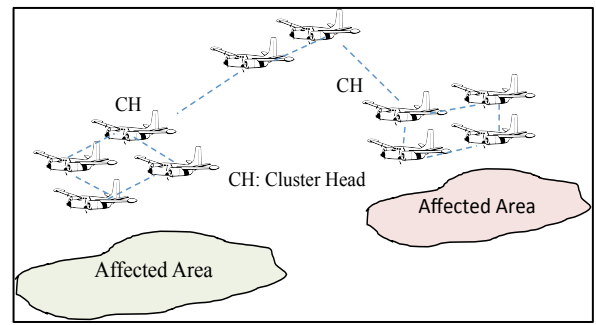

Fig. 4: Multi-Level Hierarchical Routing in UAV networks

- *Data Centric Routing:* Routing is done based on the content of data. This can be used for one-to-many

transmission in UAV networks when the data is requested by a number of nodes, It works well with cluster topologies where cluster head is responsible for disseminating information to other nodes in the cluster [3].

2) *Proactive routing protocols*

Proactive routing protocols (PRP) use tables in their nodes to store all the routing information of other nodes or nodes of a specific region in the network. The tables need to be updated when topology changes. The main advantage of proactive routing is that it contains the latest information of the routes. But to keep the tables up-to-date a number of messages need to be exchanged between nodes. This makes them unsuitable for UAV network because of bandwidth constraints. Another issue that makes them unsuitable for UAV networks is their slow reaction to topology changes causing delays [7]. Two main PRPs that have been used in MANETS/VANETS are Optimized Link State Routing (OLSR) and Destination-Sequenced Distance Vector (DSDV) protocols.

- *Optimized Link State Routing (OLSR)* is currently one of the most employed routing algorithms for ad hoc networks. Routes to all destinations are determined at startup and then maintained by an update process. Nodes exchange topology information with other nodes of the network regularly by broadcasting the link-state costs of its neighboring nodes to other nodes using a flooding strategy. Other nodes update their view of the network by choosing the next hop by applying shortest path algorithm to all destinations. OLSR [124], therefore, tracks network topology. In UAV the node location and interconnecting links change rapidly and this would cause increased number of control messages to be exchanged. Increased overhead of topology control messages would lead not only to contention and packet loss but would also put a strain on already constrained bandwidth of UAV networks. Optimization involves selecting some nodes as *multipoint relays* (MPR), which alone forward the control traffic reducing the transmission required. MPRs declare link state information to all the nodes that selected them as an MPR. MPRs also periodically announce to the network that it has reachability to all the nodes that selected it as an MPR. OLSR link-quality extension can be used to take into account the quality of links. The details are given in RFC 3636 [35] and an evaluation on several metrics in [36]. While flooding of control signals is avoided with addition of MPRs, recalculation of routes is still an issue with limited computing power of UAV nodes. *GSR, global state routing,* is another variation of link state routing that provides improvement by restricting update messages only between intermediate nodes. The size of update message is large and intermediate nodes keep changing in aerial networks. This increases both overhead and bandwidth required. *FSR, fisheye state routing*, attempts to reduce overhead by more frequent small updates to the closer "within fisheye scope" nodes. In networks where the nodes are mobile this may introduce inaccuracy. Another variation called STAR, *source-tree adaptive routing,* reduces the overheads by making the update dissemination conditional rather than periodic. However, it increases memory and processing overheads in large and mobile networks. DREAM, *distance routing effect algorithm for mobility*, has a different approach, as each node knows its geographical coordinates using a GPS. These coordinates are periodically exchanged between each node and stored in a routing table (called a location table). This consumes lesser bandwidth than link state. It is more scalable. Frequency of updates can be made proportional to velocity of nodes to reduce overhead. The maximum velocity is used to calculate the possible distance that the destination could move. This process is repeated at each hop with an undefined recovery mechanism if there are no one-hop neighbors within the wedge [125]. However, the authors in [126] have found that the complexity of DREAM does not appear to provide benefits over a simple flood.

- *Destination Sequenced Distance Vector (DSDV)* is a table-driven proactive routing protocol, which mainly uses the Bellman–Ford algorithm with small adjustments for ad hoc networks. It uses two types of update packets-"full-dump" and "incremental". Whenever the topology of the network changes, "incremental-dump" packets are sent. This reduces overhead but overhead is still large due to periodic updates. DSDV uses sequence numbers to determine freshness of the routes and avoid loops. True to the characteristics of proactive protocols, this protocol needs large network bandwidth for update procedures. Along with this the computing and storage burden of a proactive protocol puts DSDV at a disadvantage for aerial networks. In [127] authors have compared performance of DSDV with other protocols.

- BABEL is based on distance-vector routing protocol. It is suitable for unstable networks as it limits frequency and duration of events such as routing loops and black holes during re-convergence. Even when mobility is detected it quickly converges to a loop free, not necessarily optimal, configuration. BABEL, explained in RFC 6126 [37] has provisions for link quality estimation. It can implement shortest-path routing or use a metric. One of its downside is that it relies on periodic routing table updates and generates more traffic than protocols that send updates when the network topology changes [32]. In [128] Rosati et al. have shown on that based on datagram loss rate and average outage time, BABEL fails to deliver in the UAV environment.

- Better Approach to Mobile Ad Hoc Network (B.A.T.M.A.N.) [38] is a relatively new proactive routing protocol for Wireless Ad hoc Mesh Networks that could be used in environments like Mobile Ad hoc Networks (MANETs). The protocol proactively maintains information about the existence of all nodes in the mesh that are accessible via single-hop or multi-hop communication links. For each destination the next hop neighbor is identified which is used to communicate with this destination. B.A.T.M.A.N algorithm only cares about the best next-hop for each destination. It does not calculate the complete route, which makes a very fast and efficient implementation possible. Thus communication links may have varying quality in terms of packet loss, data rates, and interference. New links may appear and known links disappear frequently. B.A.T.M.A.N. considers these challenges by doing statistical analysis of protocol packet loss and propagation speed and does not depend on state or topology information from other nodes. Routing decisions are based on the knowledge about the existence or lack of information. As nodes continuously broadcast origin messages (OGMs); without packet loss, these messages would overwhelm the network. The scalability of B.A.T.M.A.N. depends on packet loss and thus it is unable to operate in reliable wired networks. B.A.T.M.A.N. protocol packets contain only a very limited amount of information and are therefore very small. If some protocol packets are lost, the information about them helps in better routing decisions. B.A.T.M.A.N.'s crucial point is the decentralization of the knowledge about the best route through the network — no single node has all the data. A network of collective intelligence is created. This approach has shown in practice that it is reliable and loop-free [39]. The self-interference caused by data traffic leads to oscillations in the network. The batman-adv performs better than B.A.T.M.A.N [129]. Comparative studies of B.A.T.M.A.N. have shown varying results. Some studies show performance of B.A.T.M.A.N. to be same as OLSR. When maximum number of nodes with maximum packet length scenario is taken, OLSR performed better than B.A.T.M.A.N. With mobility factor also OLSR performed better than B.A.T.M.A.N.. In bandwidth-limited networks all protocols gave similar throughputs but without bandwidth restrictions BABEL and B.A.T.M.A.N. perform better than OLSR for large number of users (15 or more). For small network sizes (5 mesh nodes), the three routing protocols behave similarly [40], [41]. In another study with a static wireless network of 7x7 grid of nodes, B.A.T.M.A.N. outperforms OLSR on almost all performance metrics. In another study B.A.T.M.A.N. performed 15% better than OLSR [42].

*3) Reactive routing protocols*

Reactive Routing Protocol (RRP) is an on-demand routing protocol in which a route between a pair of nodes is stored when there is communication between them. RRP is designed to overcome the overhead problem of proactive routing protocols. Because of on-demand nature, there is no periodic messaging making RRP bandwidth efficient. On the other hand, the procedure of finding routes can take a long time; therefore, high latency may appear during the route finding process. Reactive protocols can be of two types: source routing and hop-by-hop routing. In source routing each data packet carries the complete source to destination address so intermediate nodes can forward packets based on this information. No periodic beaconing is required to maintain connectivity. This does not scale well as route failure probability increases with the network size and also the header size grows increasing the overhead. In hop-by-hop routing, each data packet only carries the destination address and the next hop address. Intermediate nodes maintain routing table to forward data. The advantage of this strategy is that routes are adaptable to the dynamically changing environment. The disadvantage of this strategy is that each intermediate node must store and maintain routing information for each active route and each node may require being aware of their surrounding neighbors through the use of beaconing messages [8], [32]. Two commonly used RRPs are Dynamic Source Routing (DSR) and AODV.

- *Dynamic Source Routing (DSR)* has been designed mainly for multi-hop wireless mesh ad hoc networks of mobile nodes. It allows networks to be self-organizing and self-configuring without the need for any existing network infrastructure. DSR works entirely on demand and it scales automatically to what is needed to react to changes in the routes currently in use. It's "route discovery" and "route maintenance" mechanisms allow nodes to discover and maintain routes to arbitrary destinations. It allows selection from multiple routes to any destination and this feature can be used for applications like load balancing. It guarantees loop-free routing [43]. When applied to UAV networks, finding a new route with DSR can be cumbersome [44]. Each packet has to carry addresses of all nodes from source to destination making it unsuitable for large networks and also for networks where topology is fluid.

- *Ad hoc On Demand Distance Vector (AODV)* is a hop-by-hop reactive routing protocol for mobile ad hoc networks. It adjusts well to dynamic link conditions, has low processing and memory overhead, low network utilization, and determines unicast routes to destinations within the ad hoc network [45]. It is similar to DSR but unlike DSR each packet has only destination address so overheads are lower. The route reply in DSR carries address of every node while in AODV only the destination IP address. In AODV, the source node (and also other relay nodes) stores the next-hop information corresponding to each data transmission. There is minimal routing traffic in the network since routes are built on demand. It has been studied for possible adaptation in UAV networks. However, there are delays during route construction and link or node failures may trigger route discovery that introduces extra delays and consumes more bandwidth as the size of the network increases. The throughput drops dramatically as intermittent links become more pervasive. The number of RREQ messages increases at first, as the route discovery is triggered more frequently and after some point starts to decrease as network performance is so impacted that even RREQ messages can no longer be carried over the network [30]. Shirani et al. [139] have proposed a combined routing protocol, called the Reactive-Greedy-Reactive (RGR), for aerial applications. This protocol combines Greedy Geographic Forwarding (GGF) and reactive routing mechanisms. The proposed RGR employs location information of UAVs as well as reactive end-to-end paths

412in the routing process. Simulation results show that RGR outperforms existing protocols such as AODV in search missions in terms of delay and packet delivery ratio.

A number of researchers have compared various protocols in different settings. In one investigation, it has been found that for TCP traffic and mobile nodes, in terms of routing overheads, DSR outperforms AODV, TORA (discussed later) and OLSR as it sends the least amount of routing traffic into the network. In terms of throughput, OLSR outperformed other protocols in all the scenarios considered [46]. In another investigation effect of some modifications in AODV has been studied. AODV has a flat structure and has the potential to support dynamic networks. To take care of the significant overhead due to frequent topology changes and intermittent links, the researchers utilized some stable links and limited duplicate flooding during route construction. This scenario is possible in UAV networks because UAV can hover over a location and form relatively stable links with each other. The flat structure of AODV is then converted into hierarchical routing structure by segmenting the network into clusters. The advantage here is that duplicate route requests are limited to certain clusters instead of flooding the entire network [30].

*4) Hybrid routing protocols*

By using Hybrid Routing Protocols (HRP), the large latency of the initial route discovery process in reactive routing protocols can be decreased and the overhead of control messages in proactive routing protocols can be reduced. It is especially suitable for large networks, and a network is divided into a number of zones where intra-zone routing is performed with the proactive approach while inter-zone routing is done using the reactive approach. Hybrid routing adjusts strategy according to network characteristics and is useful for MANETs. However, in MANETs and UAV networks dynamic node and link behavior makes it difficult to obtain and maintain information. This makes adjusting routing strategies hard to implement.

- *Zone Routing Protocol (ZRP)* [47] is based on the concept of zones. Zones are formed by sets of nodes within a predefined area. In MANETs, the largest part of the whole traffic is directed to nearby nodes. Intra-zone routing uses proactive approach to maintain routes. The inter-zone routing is responsible for sending data packets to outside of the zone. It uses reactive approach to maintain routes. Knowledge of the routing zone topology is leveraged by the ZRP to improve the efficiency of a globally reactive route query/reply mechanism. The proactive maintenance of routing zones also helps improve the quality of discovered routes, by making them more robust to changes in network topology.

  Nodes belonging to different subnets must send their communication to a subnet that is common to both nodes. This may congest parts of the network. The most dominant parameter influencing the efficiency of ZRP is the zone radius. However, the cost of ZRP is increasing complexity, and in the cases where ZRP performs only slightly better than the pure protocol components, one can speculate whether the cost of added complexity outweighs the performance improvement

- *Temporarily Ordered Routing Algorithm (TORA)* [48] is a hybrid distributed routing protocol for multi-hop networks, in which routers only maintain information about adjacent routers. Its aim is to limit the propagation of control message in the highly dynamic mobile computing environment, by minimizing the reactions to topological changes. It builds and maintains a Directed Acyclic Graph (DAG) from the source node to the destination. TORA does not use a shortest path solution, and longer routes are often used to reduce network overhead. It is preferred for quickly finding new routes in case of broken links and for increasing adaptability [7]. The basic, underlying algorithm is neither distance-vector nor link-state; it is a type of a link-reversal algorithm. The protocol builds a loop-free, multipath routing structure that is used for forwarding traffic to a given destination. Allows a mix of source and the destination initiated routing simultaneously for different destinations. TORA may produce temporary invalid routes.

*5) Geographic 2-dimension and 3-dimension protocols*

A number of routing schemes have been proposed for 2-dimensional networks that can be modeled using planar geometry. Many prominent ones have been described in this survey and some have been shown to have good performance on metrics like delivery, latency and throughput in simulations. Geographic schemes assume knowledge of geographic position of the nodes. They assume that the source knows the geographic position of the node and sends message to the destination co-ordinates without route discovery. The most common technique used in geographic routing is greedy forwarding in which each node forwards the message to a node closest to the destination based only on the local information. A situation may arise when the message gets trapped in the local minimum and progress stops. The recovery mechanism is usually face routing which finds a path to another node, where greedy forwarding can be resumed [130].

In many applications, it may be more appropriate to model UAV networks in 3 dimensions. Some protocols have been proposed which, like their 2-dimensional counterparts, use greedy routing that attempts to deliver packet to the node closest to the destination. But the recovery from local minima becomes more challenging, as the faces surrounding the network hole now expand in two dimensions, and are much harder to capture. They differ mainly in the recovery methods when packets get stuck in the local minima [131].

*Greedy-Hull-Greedy (GHG)* protocol proposed in [132] involves routing on the hull to escape local minima. This is a 3 dimensional equivalent of routing on the face for 2-dimensional protocols. The authors have proposed PUDT (Partial Unit Delaunay Triangulation) protocol to divide the network space into a number of closed sub-spaces to limit the local recovery process.

In [133] the authors propose a *Greedy-Random-Greedy (GRG)* protocol using which the message is forwarded greedily until a local minimum is encountered. To come out of local minima a randomized recovery algorithm like region-limited random walk or random walk on the surface is used. The authors present simulation results for following five recovery algorithms: random walk on the dual, random walk on the surface, random walk on the graph, bounded DFS on a spanning tree, and bounded flooding. DFS on the spanning tree has shown good performance for sparse networks while random walk approaches perform well for denser networks.

*GDSTR (Greedy Distributed Spanning Tree Routing)-3D* routing scheme described in [134] uses 2-hop neighbor information during greedy forwarding to reduce the likelihood of local minima, and aggregates 3D node coordinates using two 2D convex hulls. They have shown through simulation and testbed experiments that GDSTR-3D is able to guarantee packet delivery and achieve hop stretch close to 1.

Lam and Qian contend that GHG assumes a unit-ball graph and requires accurate location information, which are both unrealistic assumptions [131]. GRG uses random recovery, which is inefficient and does not guarantee delivery. The authors propose MDT (multi-hop Delaunay triangulation) protocol, which provides guaranteed delivery for general connectivity graphs in 3 dimensions, efficient forwarding of packets from local minima and low routing stretch. The authors also provide simulation results that show that MDT provides better routing stretch compared to a number of 2-dimensional and 3-dimensional protocols. They also show its suitability for dynamic topologies with changes in the number of nodes and links.

Table VI summarizes many of the routing protocols discussed above and their applicability to the UAV environment.

### C. Routing in networks prone to delays and disruptions

According to the Delay Tolerant Networking Research Group, the term Delay Tolerant Networking relates to extreme and performance-challenged environments where continuous end-to-end connectivity cannot be assumed. Initially defined for interplanetary communication, it is now concerned with interconnecting highly heterogeneous networks together. When disaster strikes, and normal communication breaks down, lack of information flow could cause delay in rescue and recovery operations. In such situations a UAV network with delay tolerant features is considered to be one of the most effective communication methods [49]. In the absence of these features, situations like intermittent links, dynamic network, node failures and node replacements could result in breakdown of communication. Applications in such cases must tolerate delays beyond those introduced by conventional IP forwarding, and these networks are referred to as delay/disruption tolerant networks.

In many harsh situations connectivity is intermittent and networks get partitioned for long durations. This causes transmission delays longer than threshold limits described by TCP protocol and packets whose destinations cannot be reached are usually dropped, making TCP inefficient.

TABLE VI
APPLICABILITY OF PROTOCOLS TO UAV NETWORKS

| Protocol type | Problems in application to UAV networks |
| --- | --- |
| Static | Fixed tables, not suitable for dynamic topology, does not handle changes well, not scalable, higher possibility of human errors |
| LCAD | Delivery delays |
| MLHR | CH becomes single point of failure, capacity issues at CH |
| Data Centric | Network overload due to query-response. Problems as in cluster based. |
| Proactive | Large overhead for maintaining tables up-to-date, bandwidth constrained networks cannot use them, slow reaction to topology changes results in delays |
| OLSR, GSR, FSR | Higher overheads, routing loops |
| DSDV | Consumes large network bandwidth, higher overheads, periodic updates |
| BABEL | Higher overheads and more bandwidth requirement due to periodic updates |
| B.A.T.M.A.N. | Depends on packet loss, does not perform well if network is reliable. |
| On-demand or reactive | High latency in route finding. Source routing does not scale well, for large network overhead may increase because of large header size. For hop-by-hop intermediate node must have the routing table. |
| DSR | Complete route address from source to destination, scaling is a problem, dynamic network is a problem |
| AODV | Lower overhead at the cost of delays during route construction. Link failure may trigger route discovery- more delays and higher bandwidth as the size of the network increases |
| Hybrid | Hard to implement for dynamic networks |
| ZRP | Inter zone traffic may congest. Radius is an important factor may be difficult to maintain in UAV networks. Complexity higher |
| TORA | May produce temporarily invalid results |
| Geographic 3D | |
| GHG | Requires location information which may become unrealistic in many applications |
| GRG | Uses random walk recovery which is inefficient and does not guarantee delivery of messages |
| GDSTR-3D | Assumes static topology |
| MDT | None documented |

UDP provides no reliable service and cannot "hold and forward." The traditional routing protocols such AODV and OLSR also do not work properly during disruptions. In these cases, when packets arrive and there are no end-to-end paths for their destinations, the packets are simply dropped [50]. While traditional solutions do not guarantee connectivity, the protocol used in UAV networks should in some way allow buffering and forwarding of packets. The existing protocols developed for infrastructure based Internet are not able to handle data transmission in such networks and new routing protocols and algorithms should be developed to handle transmission efficiently [57]. In terrestrial networks, with mobile nodes, experiments have shown 80% delivery rate in the Disaster Information System when delay tolerance features were used [49], [52].

For UAV networks prone to intermittent links and partitioning we would need to build in tolerance to delays and disruptions. End-to-end reliability methods like multi-step request-response, acknowledgements and timed-out transmission will not work due to the long delays and disconnections in these networks. To achieve tolerance to delays and disruptions the architecture has to be based on the "store-carry-forward" (SCF) protocol in which a node stores and carries the data (often for extended periods) till it can duplicate it in one or more nearby node(s). In such a case, the node must have adequate buffering capacity to store the data until it gets an opportunity to forward it. To be able to deliver messages efficiently to their destinations, for each message the best node and time to forward data should be known. If a message cannot be delivered immediately due to network partition, then the nodes chosen to carry the message are those that have highest probability of successfully delivering the message. In SCF routing the message is moved from source to the destination one hop at a time. Selection of the path from source to destination depends on whether the topology that evolves over time is deterministic or probabilistic. If the nodes know nothing about the network states then the best they can do is to randomly forward packets to their neighbors [51], [53].

Since end-to-end paths are not available in UAV networks, traditional routing protocols that assume end-to-end path before communication starts fail to deliver. Consequently, many researchers have advanced new routing algorithms such as Direct Delivery [135], which proposes single copy method for intermittently connected networks. Nodes are assigned probability of delivery of a packet to the destination. They incorporate a hybrid scheme consisting of random and utility-based strategies. The authors claim that as compared to Epidemic method, which essentially involves multiple copies, the bandwidth required is less. Epidemic routing involves multiple copies to be sent [136]. A controlled replication algorithm called Spray and Wait has been proposed in [137].

Since UAVs are location aware due to navigational requirements, this could be used to perform geographical routing if the position of the destination is known. To conserve bandwidth and to make it possible to use routing algorithms in energy-constrained systems, viability of beacon-less geographical routing in opportunistic intermittently connected mobile networks has been explored. Beacons are regularly transmitted special messages that are used by many routing algorithms to determine a node's neighbors. They consume bandwidth and energy and the information may not be accurate.

To apply this to UAV networks, it must be remembered that resources at the UAV-nodes such as storage, power and bandwidth are limited. The delivery rate is affected by delivery distance, number of encounters, network condition and the node's movements. Delivery time is also important and in this regard it may be noticed that this architecture is not suitable for real-time contents as delays could extend to hours and days. Also, a route that is suitable currently may not remain stable for long. How long it will remain stable will depend on how fast the topology changes. If it is determined that some routes are stable then route caching can be employed to avoid unnecessary routing protocol exchanges [54]. Some of the schemes use knowledge of location of the nodes, others may flood the network with packets and yet others may use 'carrier' nodes to carry data. Let us consider briefly three types of routing methods that can be examined for UAV networks:

1. Deterministic routing
2. Stochastic routing
    a. Epidemic routing based approach
    b. Estimation based approach
    c. Node movement control based approach
    d. Coding based approach
3. Social networks based approach

According to Cardei, Liu and Wu [138], routes in networks with sporadic node contacts consist of a sequence of time-dependent communication opportunities, called contacts, during which messages are transferred from the source towards the destination. Contacts are described by capacity, direction, the two endpoints, and temporal properties such as begin/end time, and latency. Graphs at different points of time need to be overlapped to find an end-to-end path. In UAV networks, the contacts of a node keep changing over time, as their duration over which they can communicate is limited. With this in view, let us now see the suitability of various types of protocols for UAV networks.

*1) Deterministic Routing*

In deterministic routing methods, we assume that the future movement and links are completely known. In the context of UAV networks this would be possible in applications where UAVs fly in controlled formations or in applications where they have to hover over an area. If all the hosts have global knowledge of the availability and motion of the other hosts then a tree approach can be used for selecting paths. A tree is built taking source node as the root and adding child nodes and the time associated with these nodes. A final path can be selected from the tree by choosing the earliest time to reach the destination node. If the characteristic profiles are initially unknown to the hosts then they learn these by exchanging the available profiles with the neighbors. Paths are selected based on this partial knowledge. This method can be improved by requiring the hosts to record the paths that past messages have taken. The deterministic algorithms presume global knowledge of nodes and links in space and time. Handorean et al. present the Global Oracular Algorithm [56] where a host has full knowledge of characteristic profile of all hosts. The authors in [57] provide a survey of such protocols. These methods would, therefore, not be appropriate for cases where topology of the network changes frequently or availability of nodes and links is not certain. The mobility of the nodes does mean that the network topology will constantly change and that nodes constantly come in contact with new nodes and leave the communication range of others. If nodes can estimate likely meeting times or meeting frequencies, we have a network with predicted contacts [138]. In general, deterministic techniques are based on formulating models for time-dependent graphs and finding a space-time shortest path in delay prone networks by converting the routing problem to classic graph theory or by using optimization techniques for end-to-end delivery metrics.

Deterministic routing protocols use single-copy unicast for messages in transit and provide good performance with less resource usage than stochastic routing techniques. This is true in applications where node trajectory is coordinated or can be predicted with accuracy, as in interplanetary networking. A multigraph is defined where vertices represent the delay tolerant nodes and edges describe the time-varying link capacity between nodes. Deterministic routing mechanisms are appropriate only for scenarios where networks exhibit predictable topologies.

*2) Stochastic routing*

This is the case when the network behavior is random and unknown. In this situation it becomes important to decide where and when to forward the packets. One possible method to route messages in such a case is to forward them to any contacts within range. The decision could also be based on historical data, mobility patterns and other information. The protocols in this category maintain a time varying network topology that is updated whenever nodes encounter each other. For example, in Shortest Expected Path Routing (SEPR) nodes maintain a stochastic model of the network. Each node constructs a time varying graph comprising nodes they have encountered and links that reflect the connection probability between nodes. The key limitation of SEPR is that it has poor scalability due to its reliance on the Dijkstra's algorithm which runs in time of the order of $O(n^2)$, where n is the number of nodes. Moreover, it is not suitable for networks with delay tolerant features and high node mobility [58]. A routing protocol called Resource Allocation Protocol for Intentional Delay Tolerant Networks (RAPID) has been proposed that considers network resources such as bandwidth and storage when optimizing a given route metric. This is especially critical when nodes have resource constraints [59].

In UAV networks with intermittent links and opportunistic contacts, routing is challenging since the time when the nodes will come in contact and for how long are not known. When the contact does happen, it needs to be determined whether the peer in contact is likely to take the packet closer to the destination. The decision to hand over the packet to the contact depends on, besides the probability of the contact taking the packet closer to the destination, available buffer spaces in the two nodes and relative priority to forward this packet compared to other packets the node holds. Additionally, if the nodes have information about their location this information can be used to advantage. According to [138] in the situation described above, the main objective of routing is to maximize the probability of delivery at the destination while minimizing the end-to-end delay. We will discuss below some of the stochastic protocols viz. Epidemic Routing, Spray and Wait, Node movement control and Coding based.

*a) Epidemic Routing-Based Approach*

This method is used in networks of mobile nodes that are mostly disconnected [138]. This is a stochastic, flooding protocol. Nodes make a number of copies of messages and forward them to other nodes called relays. The relays transfer the messages to other nodes when they come in contact with them. No prediction is made regarding the link or path forwarding probability. In this way, messages are quickly distributed through the connected portions of the network. However, upon contact with each other nodes exchange only the data they do not have in their memory buffer. Hence it is robust to node or network failure, guaranteeing that upon sufficient number of random exchanges, all nodes will eventually receive all messages in minimum amount of time [54]. The main issue with epidemic routing is that messages are flooded in the whole network to reach just one destination. This creates contention for buffer space and transmission time

Epidemic routing uses node movement to spread messages during contacts. With large buffers, long contacts or a low network load, epidemic routing is very effective and provides minimum delay and high success rate, as messages reach the destination on multiple paths. When no information is available about the movement of nodes, packets received by a node are forwarded to all or some of the node's neighbors except from where it came.

This approach does not require information about the network connectivity. However, it requires large amounts of buffer space, bandwidth and power. Spray and wait is a variation in which each message has a fixed number of copies [137]. With this approach, the source node begins with L copies for each message. When a source or relay node with more than one copies comes in contact with another node with no copies it may transfer all n copies in normal mode or n/2 copies in binary mode (Spray phase). So after the contact both nodes have n/2 copies. With one copy a direct contact with destination is required (Wait phase). A message will be physically stored and transmitted just once even when a transfer may virtually involve multiple copies.

Approaches using indiscriminate or controlled flooding use varying amount of buffer space, bandwidth and power. There are variations like Prophet and MaxProp protocols, in which data is arranged in order of priority based on some criteria. However, these methods deliver good performance in a regular movement case. MaxProp protocol puts a priority order on the queue of packets. It determines the messages to be transmitted or dropped first. The priorities are based on the path likelihoods to peers [60] [61]. However, our airborne network topology may change without any known pattern. Also, even the random walk model may not describe movement of nodes due to horizontal as well as vertical motions. In such cases a three-dimensional architecture may better define the randomly violent topological changes and a usable path can be generated in an unpredictable moment.

In ant colony based routing strategy [140], the authors have proposed an exploration-exploitation model for UAV networks that is based on ant foraging behavior. Exploration is defined as the ability to explore diversified routes while Exploitation is the process to focus on a promising group of solutions. In networks with frequent partitioning new routes need to be explored and every contact opportunity is used to forward the messages. The authors have shown that performance comes close to Epidemic Routing (which provides an upper bound) in terms of delivery.

*b) Estimation Based Approach*

Instead of forwarding messages to neighbors indiscriminately, intermediate nodes estimate the probability of each outgoing link eventually reaching the destination. Based on this estimation, the intermediate nodes decide whether to store the packet and wait for a better chance, or decide when and to which node to forward. Variations in this could be decisions based on estimation of the next hop forwarding probability or decision based on average end-to-end metrics such as shortest path or delay [57].

A representative routing protocol for networks with delay tolerant feature is one that uses delivery estimation. PROPHET, (Probabilistic Routing Protocol using History of Encounters and Transitivity), in [141] is such a protocol. PROPHET works on the realistic premise that node mobility is not truly random. The authors assume that nodes in a DTN tend to visit some locations more often than others and that node pairs that have had repeated contacts in the past are more likely to have contacts in the future.

*c) Node-Movement-Control Based Approaches*

Nodes could either wait for reconnection opportunity with another node passively or seek another node proactively. In the reactive case, where the node waits for reconnection, there may be long unacceptable transmission delays for some applications. In the proactive case, a number of approaches have been proposed to control node mobility for reducing delays. In one class of methods, trajectories of some nodes are altered to improve some system performance metrics like delay [62]. In message ferrying approach special ferry nodes carry data either on preplanned routes with other nodes coming close to a ferry and communicating with it or ferry nodes move randomly and other nodes send a service request as a response to which the chosen ferry node will come close to the requesting node [63]. Using DataMules, nodes that move randomly, have large storage capacity and renewable energy source, is another method in this category. According to [138], controlled ferrying can be used in UAV networks in disaster recovery where UAVs can be equipped with communication devices capable of storing a large number of messages and can be commanded to follow a trajectory that interconnects disconnected partitions.

*d) Coding Based Approaches*

The concept of network coding comes from information theory and can be applied in routing to further improve system throughput. Instead of simply forwarding packets, intermediate nodes combine some of the packets received so far and send them out as a new single packet to maximize the information flow [57]. Erasure coding involves more processing and hence requires more power but improve the worst-case delay [142]. The basic idea of erasure coding is to encode an original message into a large number of coding blocks. Suppose the original message contains $k$ blocks. Using erasure coding, the message is encoded into $n$ ($n > k$) blocks such that if $k$ or more of the $n$ blocks are received, the original message can be successfully decoded. In [143], the authors have shown that network coding performs better than probabilistic forwarding in disruption prone networks. Erasure coding includes redundant data and is useful where retransmission is impractical. One aspect to be kept in mind is that in UAV retransmission would be a different path with possibly different set of nodes and also storage needed for redundant information is limited.

Hybrid Erasure Coding combines erasure coding and aggressive forwarding mechanism to send all packets in sequence during node contact. If the node battery has been drained or node loses mobility due to fault, it cannot deliver data to the destination creating a 'black hole' information loss problem. HEC solves the black hole problem by using the nodes' contact period efficiently.

*3) Social Networks Based Routing*

When mobility of nodes is not totally random and the nodes are likely to be present with greater probability in some known places then the random mobility model, used in a large number of protocols, is not realistic. The nodes that visit these places have a contact probability and will have more success in delivering bundles. Notion of groups, communities and popularity of nodes can also be exploited [31]. However, protocols may overuse the popular nodes and performance may degrade because of limited capacity of these nodes. The expected delay may also increase because of contention. Integration of some level of randomness may benefit performance. This is relatively a new area and more work needs to be done before it can be appropriately exploited [64]. Some applications of UAV networks, e.g., providing communication over a disaster prone area or communication over an oil drilling platform, may provide an environment where this type of routing can be applied.

Table VII gives DTN routing protocols and considerations for their application in UAV networks.

## IV. SEAMLESS HANDOVER

The UAV mesh nodes may be stationed over a disaster struck area to provide communication services across the target area and form a network with slow dynamic links. On the other hand, in applications like crop survey, which require a sweeping coverage of an area, UAVs may move around at required speeds. During a prolonged mission UAVs may periodically go out of service as they go out of power or develop faults. Their communication interfaces may also be shut down to conserve power, or one or more of the UAVs may be withdrawn, when less dense network is required. In all these cases the network needs to reconfigure and the ongoing voice, video or data sessions are required to be handed over to one of the working UAVs according to some predefined criteria. Handover allows for total continuity of network communication with only a minor increase of message latency during the handover process [65]. Subsection A elaborates the types of handoffs in UAV networks. Applicability of existing handoff schemes and new developments that can be used in UAV networks are in Subsection B. Subsection C discusses the IEEE standard media independent handover.

TABLE VII
DELAY/DISRUPTION PRONE UAV NETWORKS

| DTN Routing Algorithm | Considerations for application in UAV Networks |
|---|---|
| Deterministic | Useful when future availability and location of the nodes is known. In some UAV networks topology may change frequently and availability of nodes and links may not be certain. |
| Stochastic | |
| Epidemic based | Requires large buffer space per node, bandwidth and power. For UAV networks it must be noted that message delivery time depends on the buffer size. Comparatively delay is lower but with high-energy expenditure. |
| Estimation/Probability/ Statistical | Random methods work well for small networks but for large networks estimation result in large overheads. Maintains encounter but no location information. Changes in topology, like in UAV networks, affect convergence time. Delays are moderate at moderate energy consumption. Complexity is also moderate. Changes in topology affect convergence time. |
| Node movement based | UAVs can be made to follow a given trajectory that will connect source and destination nodes in partitioned networks. |
| Message ferrying | Location information is maintained. Large storage space required in ferries. Delays are high but energy expenditure is low. Can work with heterogeneous nodes. |
| Coding-based | Builds in redundant information so that retransmission is avoided. This aspect may be exploited in UAV networks. In UAVs retransmission would require finding a new path as disruptions are the norm. However, maintaining redundancy and aggressive forwarding takes additional bandwidth and buffer space. May provide better delivery rates than probabilistic in some settings but is inefficient if connectivity is good. |
| Social Networks | Applicable when some UAV nodes have 'more likely' locations. Such nodes must have higher buffer size and higher bandwidth links to avoid contention delays and losses |

### A. Handoffs in UAV networks

The real advantage of wireless mesh networks become apparent when self-organization is coupled with seamless handover to provide continuity of service to the users. Handover, or handoff as it is commonly called, is common in cellular networks, where mobile stations frequently move out of the coverage area of one cell tower and into that of a neighboring tower. Handovers can be hard or soft. In a hard or standard handover, the connection from the old network is broken before it is made with the new network. This would interrupt all the user sessions currently in progress at the mobile node (Figure 5a).

In case of soft or seamless handover, connection is made with the new network before breaking the connections from the old network. The original user sessions at the mobile station are maintained till the new link is up and handover action migrates the session to the new link as shown in Figure 5b.

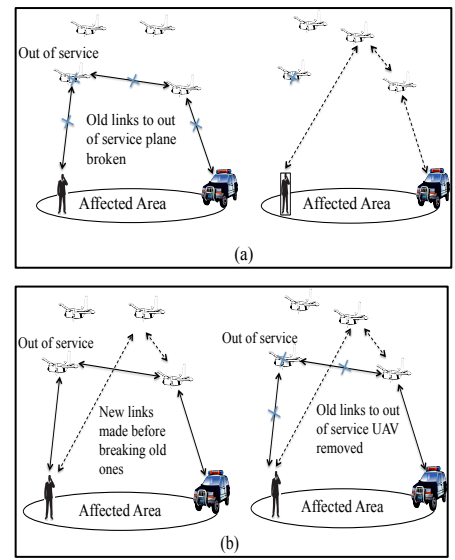

Fig. 5: a) Hard Handover b) Soft Handover

Besides being hard or soft, handovers can also be classified as *horizontal and vertical handovers*. Horizontal handovers are intra-system where the mobile access device moves from one access point to another in the same network. In the case of vertical handover, the transfer of connection is between two networks of different technologies. Figure 6 depicts these two types of handovers pictorially.

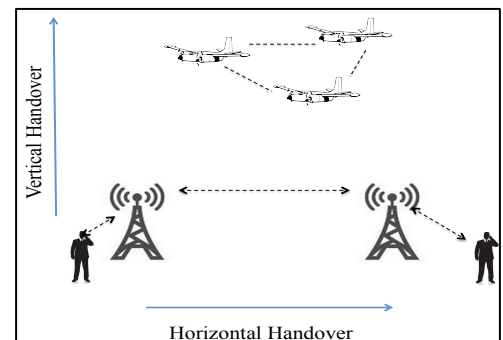

Fig. 6: Horizontal and Vertical Handovers

The handovers can also be classified based on whether the mobile access device is controlling or assisting in the handover. If both the mobile device and the network are involved then we have a hybrid handover. These have been studied in the context of VANETs and mobile IP networks but not much work is available for UAV networks [66].

### B. Applicability of Existing Handover Schemes

Lack of methods for seamless handover in UAVs, make people look at the work done in the areas of MANETS and VANETs. However, despite the need to provide seamless handover in VANETs, there are few studies regarding mobility protocols and there are no practical studies on mobility protocols using IEEE Wireless Access in Vehicular Environments (WAVE) communication [67]. WAVE consists of IEEE 802.11p and IEEE 1609.x to provide connectivity under the hostile operation conditions of VANET. However, these standards do not address the network mobility issues, either inter- or intra-technology. VANETs are characterized by the high mobility and speed of nodes, resulting in short communication time, frequent changes in network topology and network partitioning. As VANETs are a special type of MANETs, routing Protocols and IEEE standards used in MANETs have also been considered for the VANET environment. The literature survey available on vehicular communication is very limited and adaptation of the work on mobile ad hoc networks [68].

The random waypoint (RWP) model, in which mobile node movement is considered random, is commonly employed in study of MANETs. Due to the highly dynamic characteristics of VANETs, network partition or merging can occur frequently, which results in the unavailability of existing route or availability of better routes. This causes the networks to reconfigure and may trigger handovers. The handover latency and the packet loss during handover process may cause serious degradation of system performance and QoS perceived by the users. RWP model is, therefore, considered to be a very poor approximation of vehicular mobility. A more accurate and realistic vehicular mobility description at both macroscopic and microscopic levels is required [69]. The selected model should take care of temporal and spatial dependencies and geographical restrictions that the nodes would be subjected to. Two models that have been used are street random waypoint (STRAW) and Manhattan. Furthermore, the scale of VANET may vary in a large range and the node density in the network may vary dynamically depending on various application scenarios. These are important issues and more studies are required on designing handover schemes for VANET application with a view to decrease handovers and improve packet delivery and latency.

The above-mentioned considerations apply to mobile UAV networks as well. The problem is that there are only limited studies on efficient seamless handover even in IEEE 802.11-based WMNs or VANETs [67]. Many traditional mobility management protocols, such as Mobile IP version 4 (MIPv4) and Mobile IP version 6 (MIPv6) have been applied to provide handover support for VANETs and schemes for performance enhancement have been proposed. In [150] the authors propose Vehicular IP in WAVE framework along with a mobility management scheme supported by Proxy Mobile IPv6 (PIMP) over WAVE for seamless communication. They propose and analytical model that demonstrates better handover performance of VIP-WAVE as compared to standard WAVE. In order to improve MIPv6 to support the real time handover, Hierarchical Mobile IPv6 (HMIPv6) and Fast Mobile IPv6 (FMIPv6) were proposed. Inter base-station (BS) transfers of mobile devices trigger excessive signaling and reduce throughput. To counter this the Internet Engineering Task Force (IETF) proposed NEMO (network mobility). Although NEMO provides good mobility management, it does not work well in vehicular network environment [70]. As handover layer 2 and layer 3 intervention, cross layer designs that reduce packet loss (L2 function) as well as delay (L3 function) have been proposed.

In an evaluation of MIPv6 and Proxy Mobile IPv6 (PMIPv6), extended to enable enhanced support for multiple technologies, and integrated with a proposed mobility manager that monitors and selects the best access technology and network, the results show that overall the approach based on PMIPv6 performs better than the one based on MIPv6, especially when using IEEE 802.11p. Also, the former is capable of performing handover between two IEEE 802.11p networks without any data loss, even at moderate speeds [73]. The major problem in MIPv6 is the handover latency, which is the time required by vehicular node to become ready for sending or receiving packets post handover [74]. Another issue that is important is that MIPv4, MIPv6 and HMIPv6 are designed to handle terminal mobility, not for network mobility. NEMO protocol was proposed in 2005. The goal of the network mobility (NEMO) management is to effectively reduce the complexity of handover procedure and keep mobile devices connected to the Internet. The handover procedure is executed on the mobile device and mobility may cause real time services like mobile TV or voice over IP to get disconnected. Many mobility management protocols have been proposed but high degree of mobility in VANETs forces frequent handover causing problems in communication. A more rigorous evaluation of performance under various network topology and applications is required [66], [70].

In case of UAVs seamless handover procedure would involve exchange of a sequence of messages between the user mobile device and the UAV node that is being taken off service and the UAV node to which this user's session would be transferred. The handover procedure results in a transfer of physical layer connectivity and state information from one UAV to another with respect to the mobile unit in consideration. The theoretical framework, similar to that in mobile wireless networks, involves network discovery, trigger and execution. In the network discovery phase, the mobile station discovers several wireless networks on assigned channels and creates a list of APs prioritized by the received signal strength. The station may listen passively for broadcast service advertisements or actively send probes. The decision to trigger transfer is taken based on multiple criteria like signal to noise ratio. In the execution phase the actual transfer of the sessions to the new access point takes place. All contextual information related to the mobile user is transferred to the new UAV. Table VIII gives important considerations for various protocols.

### C. Media Independent Handover

IEEE 802 initially did not support handover between different types of networks. They also did not provide triggers to accelerate mobile IP based handovers. IEEE has now standardized Media independent handover (MIH) services through their standard IEEE 802.21 [73]. The key function of MIH, known as the Media Independent Handover Function (MIHF) is between the layer 2 wireless technologies and IP at layer 3 [75]. These services can be used for handovers and interoperability between IEEE-802 and non-IEEE-802 networks, e.g., cellular, 3GPP, 4G. The networks could be of

TABLE VIII
PROTOCOLS FOR SEAMLESS HANDOVER

| Protocol | Issues to be considered for UAV networks |
|---|---|
| MIPv4 | IP address shortage, weak security mechanism, auto configuration of IP [67] |
| MIPv6 | Has high handover latency due to signaling overhead, packet loss, not scalable, not efficient [71]. |
| PMIPv4 | Improves latency but does not ensure seamless handover. Performs better than MIPv6, when using IEEE 802.11p. As proxy handles signaling the mobile nodes do not need any modifications, signaling overheads lesser than MIPv4 [67]. |
| HMIPv6 | Introduced to reduce signaling between nodes and other equipment. It can also reduce handover latency due to smaller signaling and shorter path [71] |
| FMIPv6 | Improvement over MIPv6. Relies on predictive method that is difficult to make accurate for fast and randomly moving mobile nodes. It is not suitable for real time services in fast moving vehicles [71]. |
| NEMO | IETF NEMO BS based on MIPv6. It is more effective than terminal mobility. It is by itself not sufficient for seamless handover, optimization is necessary. Route optimization may not be done due to security and incompatibility issues. Latency of link layer handover and NEMO signaling overhead affect the overall performance of mobility management significantly [72]. |

*the same* or different media type, wired or wireless. This standard provides link-layer intelligence and other related network information to upper layers to optimize handovers between heterogeneous networks. It consists of signaling and triggers and makes available information from MAC/PHY to network and application layers. The standard is a hybrid implementation as it supports cooperative use of information available with the mobile station and the network. The mobile station has information about signal to noise ratios of available networks and the network knows about the information about access points, mobile stations and the higher layer service availability. The handover process can be initiated by measurement reports and triggers supplied by the link layers on the mobile station. Intra-technology handover, handover policy, security, enhancement specific to particular link layer technologies, higher layer (3 and above) enhancements are not part of this standard. MIH, however, does not provide intra-technology handover, handover policies, security and enhancements to link layer technologies [76]. Those are handled by the respective link layer technologies themselves.

In case of emergency response systems there are many jurisdictions involved. This causes inter-operability issues among jurisdictions and brings in inefficiencies in the communications. What is required is an architecture that provides a common networking platform for heterogeneous multi-operator networks, for interoperation in case of cross jurisdiction emergencies. Media Independent Handover offers a single unified interface to higher layers through technology independent primitives. It obtains information from lower layers through media specific interfaces. This allows UAV networks to interact with other technologies making it possible to use them in an integrated manner [144].

In applications like search and rescue or real-time assistance to troops in the field, using MIH, UAV networks can bring live video feeds in conjunction with other networks, viz., WiFi, WiMAX, LTE, etc. However, MIH is a nascent technology that has not been widely deployed and evaluated. In order for this to work, both mobile devices and the network must implement the standard [145].

SDN based mechanisms using OpenFlow protocols have been proposed for configuring network nodes and establishing communication paths. IEEE 802.21 Media Independent Handover procedures have been used to optimize handovers in heterogeneous wireless environment. It has also been shown how SDN can be used to support wireless communication paths with dynamic mobility management capabilities provided by standards compliant nodes. The combination allows enhanced connectivity scenarios; allowing 'always best' connectivity when multiple handover candidates are available and OpenFlow procedures are triggered preemptively in order to avoid traffic disruption [77].

## V. ENERGY EFFICIENCY IN UAV NETWORKS

There could be two scenarios in UAV networks. First, energy for communication equipment as well as for powering the UAV comes from the same source or, alternatively, they could be from different sources. In either case, energy consumption by the communication equipment is substantial and can limit the useful flying time and, possibly, the life of the network. The important point to note here is that there is a large consumption even when there is no transmission or reception, i.e., when the wireless interface is idle. Power rating of a typical Wifi 802.11n interface, in non-MIMO single antennal mode, is 1280mA/940mA/820mA/100mA under Transmission/Reception/idle/sleep modes, respectively [151]. A typical small drone may have battery capacity of 5200mAh, 11.1V. Such a drone draws about 12.5A and gives a flying time of about 25 minutes. Communication equipment of the UAV in a mesh network normally receives and transmits continuously. A quick calculation would show that the flight time would be reduced by 16%, of the rated value, if communication equipment uses the same battery as the UAV. Together with GPS and a couple of sensors, it can easily go beyond 20%. In practice, however, the net effect of this could be more severe as the battery voltage drops down below l1.1V even before the power is fully drained, inhibiting normal function of the UAV.

Let us now consider the scenario where the communication equipment has its own battery, separate from the UAV battery. In this case, the battery weight needs to be taken into account in the limited payload capacity of the UAV. In a series of experiments, the authors of this paper used separate AAA batteries to power up the airborne OpenMesh router. The router required 8 AAA batteries to provide sufficient voltage to operate. Typical alkaline batteries weigh about 11.5g each and have capacity of 860mAh. [146]. If the flight is of 25 minute duration, with continuous transmission and reception, the consumption would be 740mAh. Eight cells would theoretically work for about 9 hours but, as the voltage drops, the router would stop functioning and has to be brought down for change or recharge. In our experiments, fully charged alkaline cells provided good enough voltage for about 8 hours, enough for as many as 19 sorties. Considering the weight of the battery was another matter. Eight cells weigh 92g. The dead weight of the UAV used is about 1kg and it could carry about 300 grams of payload. The batteries constituted about 30% of the payload that could be put on the UAV! Reducing the energy consumption, therefore, would result in increase in network lifetime or increase in useful payload that can be carried.

Controlling transmission power of nodes, or making some nodes go to sleep, when network can operate without them can reduce power consumption. Nodes, that are being actively used, are carrying traffic or helping proliferate updates required by the routing protocol. The nodes that are powered up but not carrying traffic, or carrying low traffic, are still consuming energy. It is possible to avoid this waste of energy by switching off unused network elements dynamically. We shall see a number of ways this can be achieved. It must, however be mentioned that it is not just at the physical layer that the devices can be designed to save energy but also at the data link and network layers [78]. At the physical layer the built in power save modes of the devices may be exploited. At the data link layer avoiding collisions and extending sleep time in power save mode (PSM) would be important. At the network layer routing for minimizing power consumption may be the criteria. The concerns regarding power saving in the UAV networks are in some ways similar to that in mobile ad-hoc networks and wireless sensor networks. There have been a few research efforts to adapt some of the existing energy-aware protocols to the UAV networks. There are others that can potentially be used in UAV applications but are still to be tried. While there could be different ways to classify them, a commonly followed approach is to broadly classify them on the basis of the protocol layer they operate in. Within that broad classification, we also classify them on the basis of energy saving strategy used. The area is still developing therefore, rather than being exhaustive, we would aim at including the representative protocols that have been tried in some for in UAV networks or are potentially good candidates. To this end we also present some cross-layer considerations for improving energy efficiency.

It would be pertinent to mention here that the energy harvesting or scavenging alternative to power saving has been studied in recent times. Energy can be harvested from kinetic, thermal and electromagnetic sources. Creation of two queues, one for random energy arrivals and the other for data arrivals requires redesigning of transmission algorithms for wireless systems including UAV networks. This is in itself an evolving, specialized area and has not been made part of this survey.

The remaining section is divided as follows: Subsection A discusses energy efficient network layer protocols that can be used in UAV networks. Subsection B deals with data link layer protocols for conserving energy while the physical layer mechanisms are given in Subsection C.

*A. Energy Conservation in the Network Layer*

Energy efficient routing is important for networks with energy constraints. Energy efficient protocols typically use one of the standard network layer protocols as the underlying routing protocol. Many of the MANET protocols have been tried but overall the simulation results generalized in favor of one or the other. It is not clear which class of protocols would work best. Applicability of most of these to different UAV network scenarios is still to be proven. We will look here at many of the protocols that have been tried in UAV networks, as well as the potential candidates.

There are some well-known ways of building in energy conservation functionality into UAV networks. Measuring and controlling power utilization or distributing load fairly based on some energy metric are the two important ones. Additionally, some routing protocols may make some of the nodes to sleep, or navigate a low energy path, to avoid waste of energy. The protocols discussed are further classified into the following four categories: 1) Path selection based 2) Node selection based 3) Coordinator based 4) Sleep based. Various proposals for these four categories are described below and summarized at the end of this subsection.

*1) Path selection based*

These protocols aim to select paths that minimize total source to destination energy requirement. Four protocols in this category are discussed below.

*a) EMM-DSR (Extended Max-Min Dynamic Source Routing) Protocol*: This mechanism maximizes energy efficiency by finding the shortest path based on energy. It maintains a good end-to-end delay and throughput performance. It extends the Max-Min algorithm to maximize throughput, minimize delay and maximize energy efficiency. This extension has been applied to the existing on-demand dynamic source routing protocol (DSR) in the context of mobile ad hoc networks, and the resultant version takes the name of EMM-DSR [80]. However, in [55], the performance based on end-to-end latency of DSR was worst for UAV networks as compared to AODV and directional OLSR.

*b) FAR (Flow Augmentation Routing):* FAR is a transmission power optimization protocol. It assumes a static network and finds the optimal routing path for a given source–destination pair that minimizes the sum of link costs along the path. The cost depends on cost of a unit flow transmission over the link, initial and residual energy at the transmitting node. The flow augmentation algorithm requires frequent route computations and transitions but it selects the shortest cost route each time. As pseudo-stability of the topology can only be assumed in a small subset of UAV network applications (e.g. communication coverage of a remote area), there will be heavy penalty in terms of computation of minimum cost link paths when the nodes change their relative positions frequently [81].

*c) The Minimum-energy Routing:* Minimum-energy routing saves power by choosing paths through a multi-hop ad hoc network that minimize the total transmit energy. Distributing energy consumption fairly maximizes the network lifetime. In this protocol, nodes adjust their transmission power levels and select routes to optimize performance. This takes many forms in the UAV networks. Topology may be selected by adjusting the power such that only immediate neighbors communicate. It is possible to set this up in UAV networks except that the neighbors may change more frequently. The work on multi-hop communication is a swarm of UAVs [147] has reported use of minimum-energy expenditure multi-hop routing similar to ExOR (Extremely Opportunistic Routing). High power hops are split into smaller low power hops. Edge weights represent attenuation in the network and Dijkstra's algorithm is used to calculate the shortest (lowest energy) path. ExOR broadcasts each packet, choosing a receiver to forward only after learning the set of nodes that actually received the packet. ExOR operates on batches of packets in order to reduce the communication cost of agreement. The source includes in each packet a list of candidate forwarders prioritized by closeness to the destination. Highest priority node forwards the batch. Remaining forwarders forward in order but only packets that have not been acknowledged.

*d) The Pulse protocol:* A pulse, referred to as flood, is periodically sent at fixed interval originating from infrastructure access nodes and propagating through entire component of the network. This pulse updates each node about the nearest pulse source and each node tracks best route to the nearest pulse source based on some metric. The propagation of the flood forms a loop free routing tree rooted at the pulse source. If a node needs to send a packet it responds to the pulse with a reservation packet. This protocol could suffer from flood overlap delays and can result in significant consumption of energy [82]. This has been proposed in the context of multi-hop fixed infrastructure wireless networks. By analogy it can be examined for the infrastructure category of UAV applications.

2) Node Selection based

These protocols aim to select nodes that preserve battery life of each node or exclude nodes with low energy. Three such protocols are discussed below.

*a) Power-Aware Routing:* These protocols may minimize total power requirement for end-to-end communication or preserve battery life of each node. MTPR (Minimum Total Transmission Power Routing) and MBCR (Minimum Battery Cost Routing) are respectively based on these principles. In MTPR, adaptation of transmission power could result in a new hidden terminal problem. If this happens then higher number of collisions would result in more transmissions resulting in higher energy consumption. MBCR only considers the total cost function and may include a node with very less energy if all other nodes included have high energy. The variation, MMBCR (Min-Max Battery Cost Routing) takes into account the remaining energy level of individual nodes instead of the total energy. The downside is that in the quest to minimize total energy it can choose a long path and result in excessive delays [90], [118]. A variation has been proposed for wireless networks with renewable sources of energy where the node has full knowledge of the energy it will have until the next renewal point. The proposed algorithm is shown to be asymptotically optimal. The proposed routing algorithm uses a composite cost metric that includes power for transmission and reception, replenishment rate, and residual energy [83].

*b) LEAR (Localized Energy-Aware Routing): LEAR* uses DSR but allows a node to determine whether to forward the RREQ or not depending on the residual battery power being above a decided threshold value. If the residual power is below the threshold value, the node drops the message and does not participate in relaying packets. The destination node would receive a message only when all intermediate nodes along a route have good battery levels. If the source node does not receive a reply then it resends the message. Intermediate nodes lower the threshold to allow forwarding to continue [84].

*c) DEAR (Distributed Energy-Efficient Ad Hoc Routing) protocol*: Conventional power aware routing algorithms may require additional control packets for gathering information such as network topology and residual power. While it is easy to obtain such information in proactive routing, in case of on-demand protocols separate control packets may be required. DEAR uses already available RREQ packets to acquire necessary information and no extra packets would be required. DEAR only requires the average residual battery level of the entire network. Nodes with relatively larger battery energy will re-broadcast RREQ packets earlier. On-demand routing protocols drop duplicate RREQ without re-broadcasting them. DEAR can set up route composed of the nodes with relatively high battery power. Simulation shows that DEAR prolongs the network lifetime and also improves the delivery ratio by selecting a more reliable path [85].

3) Cluster head or coordinator based

In these protocols the predominant property is selection of a cluster head or a coordinator that will remain awake while the other nodes can sleep to conserve power. These protocols are quite suitable for sensor networks. They could be a natural choice in multi-level hierarchical UAV networks and in situations where multiple UAV networks need to communicate. Some examples of such protocols follow:

*a) CBRP (Cluster Based Routing Protocol):* It is an on-demand routing protocol in which nodes are divided into a number of 2-hop diameter clusters. Each cluster has a cluster head that acts as a coordinator that communicates with other cluster heads. All nodes of a cluster communicate through the coordinator. The protocol efficiently minimizes the flooding traffic during route discovery and speeds up this process as well resulting in significant energy savings [86].

*b) Distributed Gateway Selection [122]:* In this method, some superior nodes in the UAV network are selected as gateways whereas other nodes connect to the command center through these gateways. As against selection of cluster heads, selection of gateways takes into account movement of nodes. They propose an adaptive gateway selection algorithm based on dynamic network partition. The selection process goes through several iterations. The selection is based on stability of two-hop neighbors.

*c) GAF (Geographic Adaptive Fidelity) protocol:* In GAF complete network topology is divided into a grid of fixed sized squares. Nodes in two adjacent squares can communicate with each other. In each square the node with the highest residual energy is selected as the coordinator or master. Only one node in each square needs to be aware of the geographical location of other nodes in the area and this node may become the coordinator. The coordinator keeps on changing within the area. Nodes other than the coordinator can sleep without affecting routing fidelity. The protocol increases the lifetime of the network. It is a location based hierarchical protocol in which the coordinator does no aggregation [87]. In UAV networks nodes would not be stable in a grid. The connected set made out of the high energy UAVs in each square cannot persist as nodes move in an out. Frequent change of candidates for connected set in each square and presence of low energy nodes in some of the squares may lead to partitioning of the network.

*d) Span protocol:* Span reduces energy consumption without reducing connectivity or capacity. It works between the network and MAC layers and exploits the power saving feature of the MAC layer, i.e., make devices sleep when no data is to be transmitted. Each node takes a local decision whether or not to be a coordinator for when to wake up or sleep. Coordinators stay awake all the time and perform multi-hop routing while other nodes can be in power save mode. Each node gets the capability of selecting neighbors and assists in altering the network topology. The number of links is reduced and the nodes still get the advantage of using good QoS links through other nodes. In UAV networks the change of coordinators forming the backbone network would depend on the changing position of UAVs. If a number of UAVs try to become coordinators then all but one have to back off. During backoff their position may change and some may not remain candidates for forming the backbone. The amount of energy saved depends on the node density. System lifetime of an 802.11 network in power saving mode with Span is a factor of 2 better than without it [88].

4) Sleep based: These protocols conserve energy mainly by making as many nodes sleep for as long as possible. Three such protocols are described below.

*a) CDS (Connected Dominating Set) protocol:* A connected dominated node is a set of nodes such that every other node in the network is a neighbor of one of these nodes. If the members of the CDS set are connected then all other nodes will receive the packets. While dominating set is always active, other nodes can sleep to save energy [89]. With the help of CDS routing is easy and can adapt to topology changes. This protocol allows saving energy by selecting CDS nodes on energy metrics and also allowing non-CDS nodes to sleep.

*b) CaDet* (Clustering and Decision-Tree Based) protocol: This protocol uses data-mining techniques and analysis of real wireless data by probabilistic location estimation and the multiple-decision tree-based approach. These would help in identifying the most used access points that will in turn give information about the location of the users. These decisions help in selecting minimum number of access points to be used, thus reducing the power consumption and wake-up time of the client [76]. Shi and Luo [148] propose a cluster based routing protocol for UAV fleet networks. One of the criteria for choosing cluster head is residual energy. The protocol forms stable cluster architecture of UAV fleet as the basis and then performs route discovery and route maintenance by using the geographic location of UAVs.

*c) EAR (Energy Aware Routing) protocol:* In the least hops methods, with no energy considerations, the shortest path is always used and nodes on this path get depleted leading to partitioning of the network. In EAR routing and traffic engineering decisions are made taking the energy consumption of equipment into account even at the cost of sub-optimal paths. It may actually lead to multiple sub-optimal low energy paths. Energy cost can be used as the only objective or in combination with other constraints like QoS. The aim is to completely switch-off components that will help to reduce energy consumption of the network [90]. A way to achieve this by synchronizing transmit and receive times at nodes or by minimizing the necessary total received and transmit power over a route. In such a scenario, a route with more, but shorter, hops can be more energy efficient. In many UAV networks the nodes are continuously transmitting and receiving and achieving this synchronization would not be difficult [149].

Important features and energy saving strategy of network layer protocols are summarized in Table IX

TABLE IX
ENERGY CONSERVATION TECHNIQUES IN NETWORK LAYER

| Protocol | Features | Energy Saving Strategy |
|---|---|---|
| 1. Path selection based | | |
| EMM-DSR | Maximizes energy efficiency and minimizes path length. Maintains good end-to-end delay and throughput performance. | Least energy path |
| FAR | Finds minimum cost path based on initial, residual and flow of energy. | Transmit power control |
| Min Energy | Selects routes that minimize the total transmitted energy. Distributing energy consumption fairly can maximize the network lifetime. | Transmit power control |
| Pulse | Nodes connected to infrastructure send pulse flood to all the nodes. Nodes only remember best route pulse source with the lowest metric. | Least energy path |
| 2. Node selection based | | |
| Power Aware | Composite cost metric includes transmit and receive power, residual energy and replenishment rate. Good for renewable energy sources. | Optimize end-to-end power |
| LEAR | A route will be selected if nodes with high power available from source to destination for sending RREQ | Conserve energy of low residual energy nodes |
| DEAR | Only requires the average residual battery level of the entire network. Nodes with relatively larger battery energy will re-broadcast RREQ packets earlier. Prolongs network life. | Use high residual battery life routes |
| 3. Cluster-head or Coordinator based | | |
| CBRP | Each cluster head communicates with other cluster heads. Nodes maintain neighbor table. The protocol efficiently minimizes the flooding traffic and speeds up this process, reduces energy consumption. | Cluster-based, minimize flooding traffic |
| Distributed Gateway Selection | Gateways are like cluster heads. Selected based on stability | Only gateways communicate with control centers. |
| SPAN | High-energy nodes and those that add to network connectivity become coordinators. Depends on node density for energy savings. | Use only high energy nodes |
| GAF | Zoning of network. One coordinator per zone. Disadvantage is that all | Coordinator stays awake |

|  | nodes should know their geographical positions. |  |
|---|---|---|
| 4. Sleep based | | |
| CDS | Important nodes form connected set and stay active. All nodes are either part of CDS or one hop away. | Selected nodes stay awake |
| CaDet | It is probabilistic location estimation and multiple-decision tree based approach. Selects minimum number of APs to be used. | Keep nodes awake based on client density |
| EAR | Energy efficient routing decisions even at the cost of sub-optimal paths. Other constraints like QoS may be used. | Low energy paths are chosen |

*B. Energy Conservation in the Data Link Layer*

We now discuss the approaches that help conserve energy in the data link layer. It is the responsibility of the MAC sublayer of the data link Layer to ensure a fair mechanism to share access to the medium among nodes. In this process, it can play a key role in the maximization of node's energy efficiency. Power-saving mode (PSM) has been defined in 802.11, which reduces energy consumption of mobile devices by putting them in sleep mode when idle. However, a device in PSM has to be woken up to send or receive any packets. Power saving thus comes at the cost of delivery delay. IEEE 802.11e defines Automatic Power Save Delivery (APSD) [91]. IEEE 802.11n has introduced further enhancements to the APSD protocols, referred to as power save multi-poll (PSMP) [92]. As in its predecessors, there are scheduled (i.e., S-PSMP) and unscheduled (i.e., U-PSMP) versions. S-PSMP provides tighter control over the AP/station timeline by having the AP define a PSMP sequence that includes scheduled times for downlink and uplink transmissions. Wastage of energy in the MAC layer is mainly because of collisions, overheads, listening for potential traffic and overhearing traffic meant for other nodes. Protocols, therefore, attempt to remove one or more or these problems. Most of the available data link energy conservation approaches fall into the following broad categories 1) Duty cycle with single radio 2) Duty cycle with dual radio 3) Topology control 4) Cluster based. Some of the representative protocols in each of these categories are described below:

1) Duty cycle with single radio: These protocols involve use of sleep-activity schedules. However, only a single radio is used for signaling and data traffic:

*a) S-MAC (Sensor-Medium Access Control)*: It is one of the simple protocols proposed in this category. The node follows fixed cycle of sleep and active times. This will consume lesser energy than in cases where the node listen all the time when idle. The nodes make their own schedules and broadcast them to the neighbors. Cluster of nodes that know each other's schedule stay synchronized. It wastes energy handling low traffic and this will result in low throughput [93].

*b) T-MAC (Timeout-MAC)*: [42] As opposed to fixed schedules of S-MAC, this protocol uses adaptable schedules based on hearing no activity for a certain period of time. This results in improved efficiency over S-MAC [94]

*c) ECR-MAC (Efficient Cognitive Radio)*: This protocol uses multiple forwarders for each node so that multiple paths are available to the destination. The first available node is normally used, resulting in reduced wait time involved in waiting for a particular node to be free [95].

*d) SOFA (A Sleep-Optimal Fair-Attention scheduler):* this protocol uses a downlink traffic scheduler on the AP of a WLAN, called SOFA, which help its PSM clients to save energy by allowing them to sleep more, hence to reduce energy consumption. If a client still has pending packets from the beacon period but the AP is servicing other clients then this client has to remain awake till the last packet scheduled for it in the beacon period has been delivered. Therefore, a large portion of the client's energy wastage comes from the access points transmitting other clients' packets before it finishes transmitting the client's last packet to it. SOFA manages to reduce such energy wastage by maximizing the total sleep time of all clients [96], [78].

*e) MT-MAC (Multi-hop TDMA Energy-efficient Sleeping MAC) Protocol:* In order to improve the performance of energy efficiency, throughput and delay in WMSN, Multi-hop TDMA Energy-efficient Sleeping MAC (MT-MAC) protocol can be used. The main idea of MT-MAC is to divide one frame into many slots for nodes (primarily sensor) to forward data packets, with TDMA scheduling method. As collision and hidden terminal problems are considered in the scheduling algorithm, so the probability of collision in the network will be reduced and the throughput will be increased [97].

2) Duty cycle with dual radio: These protocols involve sleep-activity schedules. They use separate radios for signaling and data traffic.

*a) STEM (Sparse Topology and Energy Management)*: STEM has been one of the early CSMA-based MAC protocols for sensor networks that scheduled sleep and active periods. One of the radios is used to wake up neighboring nodes. This radio uses duty cycles. The second radio is used to send data to the forwarding node and is always asleep until woken up. As waking up involves some delay there is a tradeoff in which higher efficiency comes with higher delay [98].

*b) PTW(Pipeline Tone Wakeup) scheme*: This protocol aims to take care of the tradeoff between energy efficiency and delay present in STEM. It uses techniques to shorten the time taken to wake up the nodes for data transfer [99].

*c) LEEM (The Latency minimized Energy Efficient MAC) protocol:* This protocol also aims to reduce latency while maintaining energy efficiency. The nodes along the path to destination are synchronized so that they can be woken up sequentially [100].

*d) PAMAS (Power-aware Multi Access Protocol with Signaling):* This protocol uses signaling on a separate radio to get good throughput and latency for the data. It overhears signaling for other nodes and turns off a node's radio for the duration for which the medium will not be available. The signaling interface could be left on so that the node listens to the signal exchange and can respond fast to the traffic directed towards it [101].

3) Topology/power control based: These protocols adjust transmit power levels to control the connectivity to other nodes or select topology by deciding which nodes to keep on:

*a) XTC (eXtreme Topology Control)*: Neighbors are ranked

for link quality by each node broadcasting at maximum power. Each node transmits its ranking results to neighboring nodes. Finally, the nodes select the neighbors to be directly connected to. The protocol does not need location information [102].

b) *LFTC(Location Free Topology Control)*: LFTC protocol constructs power-efficient network topology. It also avoids any potential collision due to hidden terminal problem. Initially each node broadcasts hello message with vicinity table. The nodes then adjust their transmission power to communicate with direct neighbors. Data collision and hidden terminal problems are avoided by choosing appropriate power for data and control. This protocol improves over XTC as it has smaller hop count and better reliability [103].

c) PEM (*Power-Efficient MAC*) Protocol: PEM enhances the network utilization and reduces the energy consumption. The stations estimate their distances from the transmitter and obtain the interference relations among transmission pairs through a three-way handshaking. Transmission is scheduled based on the interference relation [104].

d) Virtualization of NICs: An obvious way to save energy is to switch off some of the wireless nodes during the off-peak hours. Of course, some of the locations in the area of interest would not be covered. These can be provided connectivity by network coverage extension/relaying capability. Not only shutting down nodes saves energy, NIC virtualization also reduces energy consumption [105].

4) Cluster based MAC protocols: These protocols divide nodes into groups and select a cluster head or a coordinator for inter-area communication and data aggregation. Non-cluster nodes can then have power saving sleep schedules:

a) LEACH (Low-Energy Adaptive Clustering Hierarchy) protocol: LEACH is a scheduled MAC protocol with clustered topology. It is a hierarchical routing algorithm that combines reduction in energy consumption with quality of media access. It divides the network topology into clusters and a cluster head represents each cluster. The cluster head aggregates data for all the nodes in a cluster and provides better performance in terms of the lifetime [106]. The independent decision of whether or not to become a cluster head may lead to accelerate energy drainage for some of the nodes in the network. LEACH-C is a centralized scheme, which can improve some of the issues related to LEACH.

b) PEGASIS (*Power-Efficient Gathering in Sensor Information Systems*): This protocol makes use of cluster heads in a different way from LEACH and is an improvement over it. Instead of each cluster head communicating with the destination, a chain of cluster heads is formed. The chain is constructed starting with the node farthest away from the destination and finally ending up closest to the destination. Then a cluster head is chosen that will perform data aggregation. In case of failure of any of the nodes, the chain is re-constructed. A new leader is selected randomly during each round of data gathering [107].

Important features and energy saving strategy of the data link layer protocols are described in Table X.

## C. Energy conservation in the physical layer

Physical Layer (PHY) is concerned with the hardware implementation, connectivity to neighbors, implements encoding and signaling and moves the bits over the physical medium. For designing energy efficient protocols it is important to consider the physical characteristics of the electronics involved. Understanding of physical layer is important for successful implementation of data-link and network layers. High mobility of nodes in UAV networks creates issues for physical layer. We will consider these protocols in the following four groups: 1) Dynamic voltage control 2) Node level power scheduling 3) Choosing minimum subset, and 4) Sleep schedules with buffering:

TABLE X
ENERGY CONSERVATION TECHNIQUES IN DATA LINK LAYER

| Protocol | Features | Energy Saving Strategy |
|---|---|---|
| 1. Duty cycle with single radio | | |
| S-MAC | Nodes have duty cycles. Nodes broadcast their schedules to neighbors. Consumes less energy than protocols that listen all the time. | Scheduled sleep times for nodes |
| T-MAC | Schedules based on activity. Has better efficiency than S-MAC | Adaptable schedules of sleep times |
| ECR-MAC | Each node has many forwarders. The first available is used to reduce wait time. | Reduced wait time for forwarding data |
| SOFA | Increases sleep time by allowing nodes not to wake up when other nodes packets are transmitted. | Sleep more |
| MT-MAC | Divides one frame into many slots for sensor nodes to forward data packets. Collisions are reduced and network throughput will be increased. | Increased sleep time |
| 2. Duty cycle with dual radio | | |
| STEM | It has scheduled sleep and active period. Separate radio for signaling and data. The signaling radio has scheduled times while data radio sleeps till woken up. Higher delay. | Data radio sleeps longer |
| PTW | Takes care of the tradeoff between energy efficiency and delay of STEM. Shortens the time taken to wake up the nodes for data transfer. | Data radio sleeps longer, pipelining used to reduce delay |
| LEEM | Reduce latency while maintaining energy efficiency. Nodes along the path to destination are synchronized so that they can be woken up sequentially | Nodes synchronized for fast wakeup |
| PAMAS | Signaling radio is left on for listening. Data radio of a node is switched on when it receives packet meant for it. | Data radio sleeps when medium is not available |
| 3. Topology/power control based | | |
| XTC | Each node connects to neighbors based on link quality. | Good quality link selection |
| LFTC | Nodes change topology by adjusting power. Data collision and hidden terminal problems are avoided. Improves over XTC on hop count and data delivery. | Transmission power control for data and control |
| PEM | Stations can estimate their distances from the transmitter and obtain the interference relation among transmission pairs. Stations schedule their transmission according to interference relation. | Transmission scheduling based on interference |
| Virtualization | Switch off some nodes and provide | Shutting node and |

| of NICs | connectivity to cut off areas by coverage extension, relaying. NIC virtualization also reduces energy consumption. | virtualization |
|---|---|---|
| 4. Cluster based | | |
| LEACH | Data aggregation by cluster head. Nodes have duty cycles. It is a hierarchical routing algorithm that combines reduction in energy consumption with quality of media access. | Cluster based, nodes communicate with cluster heads and have sleep-activity schedules |
| PEGASIS | Divides network into clusters. A path of cluster heads is formed and one of them is chosen to perform data aggregation. | Reducing cluster heads that perform forwarding |

The physical layer deals with modulation and signal coding and related signal transmission technologies. The extremely high mobility nodes of aerial networks make design of the physical layer more involved. In case of UAV networks the movement is in 3D and in order that data is not lost in data communication architectures the physical layer conditions have to be well understood and well defined. The 3D networks have several accentuated concerns, such as, variations in communication distance, direction of the communicating pairs, antenna radiation pattern, shad- owing from the UAV and onboard electronic equipment, environmental conditions, interferences, and jamming [150]. The authors suggest mitigating high link loss and variations by spatial and temporal diversity. They feel that more research is required in the physical layer and antenna propagation for 3D environments.

*1) Dynamic voltage control*

These protocols adjust circuit bias voltage levels, depending on the traffic, to optimize energy usage. In this regard as the UAV consumes battery power, the battery voltage may drop and the UAV may become dysfunctional. However, if the router, and other circuits in the payload can fallback to lower voltage levels then the flying time of the UAV can be increased. The following protocol is an example.

*a) Dynamic Voltage Scaling (DVS):* DVS exploits variability in processor workload and latency constraints and realizes this energy-quality trade-off at the circuit level. When the traffic load is low, the bias voltage can be reduced to get proportionally quadratic saving in energy [108]

*2) Node level power scheduling*

These protocols do node level optimizations to achieve energy efficiency and enhance network lifetime. In UAV controlling the transmission power can control the topology of the networks. In the extreme case connectivity can only be maintained with single hop neighbors. Communication can then take place in a multi-hop manner. The following protocols are in this category.

*a) LM-SPT (Local Minimum Shortest Path-Tree):* This method improves power efficiency by a localized distributed power-efficient topology control algorithm. The main idea in this approach is to construct an overlay graph topology over the wireless mesh such that this topology has the required features like increased throughput, increased network lifetime and maintained links by varying transmission power at each node. The algorithm balances energy efficiency and throughput in wireless mesh network without the loss of connectivity. The concept of this approach is based on information of the local neighborhood that is confined to one hop for calculating the minimum power transmission [109].

*b) Minimum-energy topology:* This method generates a minimum-energy topology graph $G = (V, E)$ where V represents nodes and E represents the links. It has a localized distributed topology control which calculates the optimal transmission power to maintain connectivity and reduces power to cover only nearest neighbors. Energy is saved and lifetime of the network also increases [110].

*c) CPLD (Complex Programmable Logic Devices)*: CPLDs are used with popular application chipsets to reduce standby power and minimize the time that key processors need to be powered on to detect system events. To minimize power consumption in the network, CPLD may be used in the nodes of the mesh network, which cuts off the power from the processor of the station while it is being idle for some specific time. It can detect any beacon frame arriving from the base station and switches the power on immediately to the processor without wasting any time. Implementation of CPLDs in devices, which are part of wireless mesh networks instead of microcontrollers, coupled with smart electronic switches may drastically reduce the power consumption [115].

*3) Choosing Minimum Subset*

These protocols work on the principle of choosing a subset of nodes to switch on and keep the other nodes switched off in order to save energy. Three examples of protocols in this category are as follows.

*a) CNN (Critical Number of Neighbors):* CNN refers to the minimum number of neighbors that should be maintained by each node in the network to be asymptotically connected. This approach to maintain connectivity only requires knowledge of the network size is required to determine the CNN. This information can be easily obtained from a proactive routing protocol such as OLSR. Power is controlled for individual nodes to maximize power savings and controlling interference [111].

*b) Virtual WLAN:* This method proposes a switching scheme that aims to power on the minimum number of devices or the combination of devices that consume the least energy that can jointly provide full coverage and enough capacity. By consolidating hardware, some hardware can be put in low-power mode and energy consumption can be reduced. Depending on the type of device, different amounts can be saved [112].

*c) Energy Savings in Wireless Mesh Networks in a Time-Variable Context:* This is an approach to minimize the energy in a time varying context by selecting dynamically a subset of mesh BSs to switch on considering coverage issues of the service area, traffic routing, as well as capacity limitations both on the access segment and the wireless backhaul links. To achieve the desired objective, the algorithm considers traffic demands for a set of time intervals and manages the energy consumption of the network with the goal of making it proportional to the load [113].

*4) Sleeping Schedules with Buffering*

These protocols aim to prolong sleep times by buffering packets that arrive while a node is inactive. An example is discussed here:

*a) SRA (Sleeping and Rate-Adaptation):* There are two forms of power management schemes that reduce the energy consumption of networks. The first puts network interfaces to sleep during short idle periods. To make it effective small amounts of buffering is introduced for sleeping clients that will store the packets to create a long enough gap for the client to sleep and save energy. Potential concerns are that buffering will add too much delay across the network and that bursts will exacerbate loss. The algorithms arrange for routers and switches to sleep in a manner that ensures the buffering delay penalty is paid only once (not per link) and that routers clear bursts so as to not amplify loss noticeably [114].

Important features and the energy saving strategy of the physical layer protocols are given in Table XI

TABLE XI
ENERGY CONSERVATION IN PHYSICAL LAYER

| Protocol | Features | Energy Saving Strategy |
|---|---|---|
| 1. Dynamic voltage control | | |
| DVS | Circuit voltage adjusted according to processor load and traffic latency requirements. | Low voltage (low power) for low loads |
| 2. Node level power scheduling | | |
| LM-SPT | Transmission power is varied at each node to get a reduced topology meeting the objectives like throughput, network lifetime or link quality. | Control power to control topology |
| Minimum energy topology | Optimal power to maintain connectivity with the nearest neighbors used. Energy is save and network lifetime increased | Node level transmit power control |
| CPLD | Adaptive power-on time control for electronics. CPLD cuts off power when node is idle for a specific time. Smart electronics | Device level power management |
| 3. Choosing minimum subset | | |
| CNN | Selects minimum number of neighbors for each node for complete connectivity. Power is controlled for individual nodes to maximize power savings and controlling interference. Requires knowledge of network size | Neighbor selection and power control for each selected node. |
| Virtual WLAN | Devices that will consume least energy and provide the required coverage and capacity are powered on. Remaining hardware is in low-power mode | Active node selection based on coverage and capacity. |
| Energy saving WMN | Energy consumption is proportional to the traffic load. Enough nodes are selected in each time interval to meet the coverage, capacity and traffic routing requirements. | Traffic demand based active node selection |
| 4. Sleep schedules with buffering | | |
| SRA | Nodes are allowed to sleep longer by storing packets. The packets are then forwarded in bursts during active windows. Burst losses are minimized. | Increase node sleep time with buffers and managing delays |

*D. Energy Conservation through Cross-Layer Protocols*

Protocols considered so far work predominantly on solving issues relating to one individual protocol layer. The network layer protocols deal with path or node selection for maintaining connectivity of the entire network. The data link layer protocols (specifically the MAC sub-layer protocols) are more involved in avoiding collisions and designing duty cycles. The physical protocol concerns with device characteristics for voltage and power control. Maximizing energy conservation in one layer may still give overall inefficient energy conservation across all layers. Very little work is available in the area of cross-layer protocols in mobile ad hoc network and application of these to UAV networks is still an open issue. We discuss here a couple of representative work in this area.

*1) MTEC (Minimum Transmission Energy Consumption) Routing protocol with ACW (Adaptive Contention Window):* To design an energy efficient protocol a number of factors like proportion of successful data transmissions, residual energy on the nodes and traffic condition on the nodes need to be considered. This cross-layer design can decrease the energy consumption of data transmissions, but it also prolongs network time. MTEC routing protocol works at the network layer and takes into account the proportion of successful data transmissions, the traffic load of nodes and the number of nodes contending for a channel, to find a suitable path. It reduces energy consumption and prolong network lifetime. For this it finds nodes with sufficient residual energy for successfully transmitting all data packets. ACW works at the MAC layer to reduce energy consumption and throughput. Based on the proportion of successful data transmissions, a node uses an ACW to dynamically adjust the back-off time between different nodes. It reduces re-transmissions by allowing links with higher proportion of successful transmissions to use channel more [116].

*2) CLEEP (Cross Layer Energy Efficient Protocol)* The protocol works across physical, MAC and network layers. At the physical layer level, this protocol first obtains the transmission power required to keep two neighboring nodes connected. It maintains a connected neighbor table for each node in the network. This information can be utilized by the network layer to choose a better routing path for data. The protocol then utilizes the routing information to decide the sleep-activity pattern of the nodes in the MAC layer for maximizing sleep times [117].

*3) $CLE^2aR^2$ (cross-layer energy-efficient and reliable routing protocol)*
This has been proposed to find an energy efficient and reliable route from the source to the destination. It counters channel quality variation and reduces the retransmission cost for wireless ad hoc net- works. For each node, cost of relaying data from the source to destination is calculated if this node falls in the route. This also takes into account retransmission cost in terms of energy consumption. The node receiving an RREQ also estimates the channel quality based on the received signal strength. The node also takes its interference pattern into account. With all these calculations CLE2aR2can find a route with less energy consumption and high reliability [119].

## VI. CONCLUSIONS

UAV networks are growing in importance and general interest for civil applications. Providing good inter-UAV connectivity and links to the users and any ground station is quite challenging. Research relating to mobile ad hoc mesh networks is being applied to the UAV networks, but even the former is an evolving area. Additionally, a number of features like dynamicity of nodes, fluid topology, intermittent links, power and bandwidth constraints set UAV networks apart from any other that have been researched before. Some researchers believe that there is need to re-build everything ground up. This includes features in the physical layer, data link layer, network layer and the transport layer. Some issues like energy conservation and ensuring adequate quality of service require cross-layer design.

In this survey we attempt to focus on research in the areas of Routing, seamless handover and energy efficiency. The reason to undertake this survey was lack of a survey focusing on these issues. In order to effectively process and present the available information in correct perspective, it was considered necessary to categorize the UAV networks based on a number of characteristics. It is important to distinguish between infrastructure and ad-hoc UAV networks, applications areas in which UAVs act as servers or as clients, star or mesh UAV networks and whether the deployment is hardened against delays and disruptions. Through this discussion we see how despite sharing some characteristics with mobile and vehicular ad-hoc networks, UAV networks have their own unique properties. Having done this classification, we focus on the main issues of routing, seamless handover and energy efficiency in UAV networks

Routing has unique requirements - finding the most efficient route, allowing the network to scale, controlling latency, ensuring reliability, taking care of mobility and ensuring the required quality of service. In UAV networks, additional requirements of dynamic topology (with node mobility in 2-D and 3-D), frequent node addition and removal, robustness to intermittent links, bandwidth and energy constraints make the design of a suitable protocol one of the most challenging tasks [3]. This area is still evolving and we are still to see native UAV network protocols. Researchers have proposed modifications to existing protocols to make them workable for UAV scenarios. We discuss four categories of protocols and see the extent to which they are suitable for UAV networks. Static protocols have limited applicability as routing tables are manually configured and cannot be changed once the network has been launched. Proactive protocols keep up-to-date tables but need to exchange a number of messages between the nodes. This makes them unsuitable for UAV networks on two counts – bandwidth constraints and slow reaction to topology changes causing delays. In this category, a well known protocol called OLSR would track fast changes in the UAV network at the cost of increased overhead of control messages leading to contention, packet loss and bandwidth consumption even with Multi Point Relays (MPR). Another traditional protocol, DSDV, when used in aerial networks, puts a high computing and storage burden for maintaining freshness of routes. Newer protocols like BABEL and B.A.T.M.A.N have also not found to be distinctly superior. In reactive protocols like DSR, finding new routes is cumbersome. AODV has been studied for possible adaptation in UAV networks but delays in route construction triggers route discovery and compounds delays. Throughput suffers due to intermittent links. Some adaptations like reactive-greedy-reactive (RGR) have been shown to perform better. Hybrid protocols present a compromise between higher latency of reactive protocols and higher overhead of proactive protocols. In some cases like the ZRP protocol, the added complexity in the UAV networks may outweigh the slight improvement in performance.

In applications where UAV networks are delay and disruption prone and partitioning is the norm rather than exception, continuous end-to-end connectivity cannot be assumed. The transmission delays increase beyond TCP threshold limits and packets are dropped. In such situations a UAV network with delay tolerant features is considered to be one of the most effective. The architecture would be based on store-carry-forward. If the data cannot be delivered immediately then the nodes chosen to carry the message are those that have highest probability of delivering the message. Selection of the path from source to destination depends on whether the topology that evolves over time is deterministic or probabilistic. In case of probabilistic topology, messages may be sent through flooding (epidemic routing) with large requirement of buffer space, bandwidth and power. Variation like Spray and Wait use fixed number of copies and reduce resource requirement. Controlled ferrying can be used in UAV networks in disaster recovery where UAVs can be equipped with communication devices capable of storing a large number of messages and can be commanded to follow a trajectory that interconnects disconnected partitions. Some applications like communication service in a remote area or over an oil rig could benefit from social network model where nodes remain in some known locations.

Seamless handover allows for total continuity of network communication with only a minor increase of message latency during the handover process. The handover latency and the packet loss during handover process may cause serious degradation of system performance and QoS perceived by the users. There has been hardly any study on seamless handover in the UAV environment and more so using IEEE Wireless Access in Vehicular Environments (WAVE) suite of protocols. Many mobility management protocols have been proposed but high degree of mobility forces frequent handover and problems in communication. IEEE has standardized Media Independent Handover (MIH) services through their standard IEEE 802.21. These services can be used for handovers and interoperability between IEEE-802 and non-IEEE-802 networks, e.g., cellular, 3GPP, 4G. MIH, however, does not provide intra-technology handover, handover policies, security and enhancements to link layer technologies. However, MIH is a nascent technology that has not been widely deployed and evaluated.

Energy efficiency is a very important requirement in UAV networks. Reducing the energy consumption helps in increase in network lifetime and useful payload that can be carried. Energy consumption can be reduced through transmission power control, load distribution or making nodes sleep. At the physical layer transmission power can be reduced to the minimum required for connectivity. Network layer can use the information about connected nodes to route packets. The data link layer schedule on/off times of signaling and data carrying radios. Cross layer protocols will offer schemes working at two or more layers.

There have been some good studies that have focused on a number of issues like applications, protocols and mobility [3], [12], [15]. In the present survey we focus on issues relating to characterization of UAV networks, routing under constraining circumstances, automating control with SDN, seamless handovers and greening of UAV networks that have not been the focus of the earlier surveys. To the best of knowledge this is the first survey that bring forth the current research in these areas and their importance in building successful multi UAV networks. We expect that this survey would spur more research work in these important but understudied areas with open research issues.

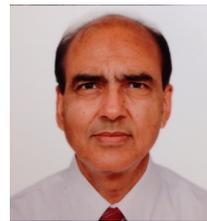

**Lav Gupta** is a senior member of IEEE. He received BS degree from Indian Institute of Technology, Roorkee, India in 1978 and MS degree from Indian Institute of Technology, Kanpur, India in 1980. He is currently pursuing PhD degree in Computer Science & Engineering at Washington University in St Louis, Missouri, USA.

He has worked for about 15 years in the area of telecommunications planning, deployment and regulation. With the sector regulatory authority he worked on technology and regulation of next generation networks. He has also worked as senior teaching faculty of Computer Science and Access Network Planning for a number of years in telecommunications academies. He is the author of one book, 5 articles and has been a speaker at many international seminars.

He was recipient of best software award from Computer Society of India in 1982 and best faculty award at Etisalat Academy, UAE in 1998.

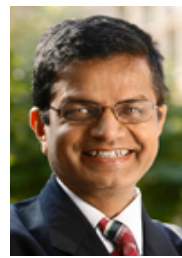

**Raj Jain** is a Fellow of IEEE, a Fellow of ACM, a Fellow of AAAS. He received BS degree in Electrical Engineering from APS University in Rewa, India in 1972 and MS in Computer Science & Controls from IISc, Bangalore, India in 1974 and the Ph.D. degree in Applied Math/Computer Science from Harvard University in 1978.

Dr. Jain is currently a Professor of Computer Science & Engineering at Washington University in St. Louis. Previously, he was one of the Co-founders of Nayna Networks, Inc - a next generation telecommunications systems company in San Jose, CA. He was a Senior Consulting Engineer at Digital Equipment Corporation in Littleton, Mass and then a professor of Computer and Information Sciences at Ohio State University in Columbus, Ohio. He has 14 patents and has written or edited 12 books, 16 book chapters, 65+ journal and magazine papers, and 10e5+ conference papers.

He is a winner of ACM SIGCOMM Test of Time award, CDAC-ACCS Foundation Award 2009, and ranks among the top 100 in CiteseerX's list of Most Cited Authors in Computer Science.

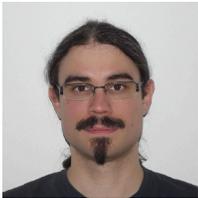

**Gabor Vaszkun** received his BS and MS degrees in Computer Science from Budapest University of Technology & Economics, Budapest, Hungary in 2012 and 2014 respectively.

Mr. Vaszkun is currently working with Ericsson Hungary. During 2014 he was a Research Scholar in the Department of Computer Science & Engineering at Washington University in St. Louis, Missouri, USA. He has interned with Ericsson Research for 3 months during his Masters program. He is the author of two published conference papers. He has also participated in a number of Poster Sessions and Demonstrations in national and International Conferences.

He received Rosztoczy scholarship for the year 2014 to pursue his research in USA. He is also a recipient of the WorldQuant scholarship awarded to outstanding students to pursue higher education in the fields of science and quantitative studies.